\documentclass[12pt,longbibliography]{article}

\usepackage[utf8]{inputenc}
\usepackage{cochineal}
\usepackage[T1]{fontenc}

\usepackage[a4paper, margin=2.5cm]{geometry}

\usepackage{amsmath, amsthm, amsfonts, amssymb}
\usepackage{mathrsfs}           
\usepackage{xcolor}
\usepackage[colorlinks=true,linkcolor=blue,citecolor=blue,urlcolor=blue,breaklinks]{hyperref}

\usepackage{bbm} 
\usepackage{multirow}

\usepackage{breakurl}
\usepackage{natbib}
\usepackage{graphicx}

\newcommand{\ts}{\textsuperscript}

\theoremstyle{definition}

\theoremstyle{remark}

\theoremstyle{example}

\title{\Large \textbf{Exploring the Effects of COVID-19 Containment Policies on Crime: An Empirical Analysis of the Short-term Aftermath in Los Angeles\\ \color{blue} \large Forthcoming at \textit{American Journal of Criminal Justice}}}
\author{Gian Maria Campedelli$^{\dagger, 1}$  \and Alberto Aziani$^{2,3}$ \and Serena Favarin$^{2,3}$}

\date{\small{$^{\dagger}$ Corresponding Author: \texttt{\href{mailto:gianmaria.campedelli@unitn.it}{gianmaria.campedelli@unitn.it}}\\
$^1$ University of Trento --- Department of Sociology and Social Research, Trento (Italy) \\
$^2$ Università Cattolica del Sacro Cuore --- Department of Political and Social Sciences,  Milan (Italy) \\
      $^3$ Transcrime - Joint Research Centre on Transnational Crime, Milan (Italy)} \\
      1\ts{st} October 2020}

\begin{document}
\newgeometry{top=0.5cm,bottom=2cm,right=2cm,left=2cm}
\maketitle

\begin{abstract}
\noindent This work investigates whether and how COVID-19 containment policies had an immediate impact on crime trends in Los Angeles. The analysis is conducted using Bayesian structural time-series and focuses on nine crime categories and on the overall crime count, daily monitored from January 1\textsuperscript{st} 2017 to March 28\textsuperscript{th} 2020. We concentrate on two post-intervention time windows—from March 4\textsuperscript{th} to March 16\textsuperscript{th} and from March 4\textsuperscript{th} to March 28\textsuperscript{th} 2020—to dynamically assess the short-term effects of mild and strict policies. In Los Angeles, overall crime has significantly decreased, as well as robbery, shoplifting, theft, and battery. No significant effect has been detected for vehicle theft, burglary, assault with a deadly weapon, intimate partner assault, and homicide. Results suggest that, in the first weeks after the interventions are put in place, social distancing impacts more directly on instrumental and less serious crimes. Policy implications are also discussed.

\end{abstract}
\normalsize
\restoregeometry
\section*{Introduction}
\label{sec:intro}
In the first months of 2020, California was one of the first States to be affected by the spread of a new virus belonging to the coronavirus family, named Sars-CoV-2. On March 4\textsuperscript{th}, six cases of COVID-19 were confirmed in Los Angeles County rising the total number of cases for the county up to seven. Following this, the Los Angeles County Board of Supervisors and the Department of Public Health declared a health emergency. From that moment on, the Los Angeles population had been invited to adopt simple social distancing strategies that limit their exposure to others—e.g., remaining home when sick—and to prepare for the possibility of more significant social distancing requirements. The institutional intervention started to become stronger on March 16\textsuperscript{th}, with the prohibition of all events comprising fifty or more attendees. On March 19th, the California Department of Public Health further reinforced the containment strategy by ordering all individuals living in the State to stay at home \citep{CountyofLosAngelesPublicHealthIssues2020}.

Distancing measures simultaneously affect the daily routines and the social interactions of millions of people. Daily commuters are forced to spend their days at home; household members share the same living spaces throughout the entire day; people can connect to their peers only telematically. Yet, the impact of the lock-down policies, and of the virus itself, outreach the alteration of people’s everyday routines and social relations. In the medium term, the losses due to the slowdown of the economic system may transform into higher unemployment and destitution for many. The scale of social distancing and lock-down policies adopted to mitigate the deadly consequences of the COVID-19 constitutes an unprecedented instrument to investigate contemporary societies and the short-term changes in crime trends.

Despite the rapidly growing attention of scientists to the consequences on crime of the COVID-19 public health emergency, examinations of the impact on different types of crimes of milder and stricter policies remain underdeveloped. Moreover, our understanding of how well certain theoretical frameworks provide support in explaining immediate changes in the occurrence of different crimes in response to the pandemic is limited. This work attempts to address these gaps by investigating the extent to which measures taken to contain COVID-19 impact on nine crime categories, and on the overall crime count, in the city of Los Angeles in the immediate aftermath of their promulgation. Specifically, we concentrate on a set of primarily instrumental crimes, namely burglary, theft, shoplifting, robbery, and vehicle theft and on more ‘expressive’ offenses, namely battery (simple assault), intimate partner assault, assault with a deadly weapon, and homicide.\footnote{We rely on the categorization provided by \cite{CohnEvencriminalstake2003}, who state that that “Expressive crime involves [...] violence that is not directed at the acquisition of anything tangible or designed to accomplish anything specific other than the violent outcome itself. Assaults, disorders, and domestic violence are examples of expressive crime. Instrumental crime [...] involves behavior that has a specific tangible goal, such as the acquisition of property. Predatory crimes, such as theft, burglary, and robbery, are examples of instrumental crime” \citep[p.~252]{CohnEvencriminalstake2003}. Some authors have argued that also assault and domestic violence–and within it, intimate partner assault–are goal-oriented because, through the commission of these crimes, offenders seek to gain control over another person or assert their identity \citep{TedeschiViolenceaggressioncoercive1994}. We agree that all crimes can have a certain extent of rationality. Still, as argued by \cite{CohnEvencriminalstake2003}, some crimes are more instrumental than others. As an example, “[a]lthough robbery is more commonly classified as a violent crime, the violence involved is usually subservient and instrumental to the goal of taking another person’s property” \citep[p.~359]{CohnEvencriminalstake2003}.}

The significance of this work consists in the importance of investigating whether and how these major societal modifications to the lives of millions of people influence the occurrence of different crimes and to reason on which criminological theories are better suited to explain crime trends during this peculiar short period. Specifically, it is relevant to understand what types of crime, if any, are most influenced by the forced modification of everyday habits and behaviors and why. We do so by exploiting the discontinuity introduced by the adoption of the social distancing requirements in Los Angeles County—more than 10 million residents—in response to the local and public health emergency caused by the spread of the COVID-19. In particular, we use a Bayesian statistical framework to derive counterfactual scenarios and estimate the causal impact of such policies, showing its potential for criminological research.

Implications of the results obtained are numerous because of the extent of the changes introduced by these policies and the novelty of the research approach proposed. The analysis is intended to verify what happened immediately after the introduction of the containment policies in order to better understand the latest evolution of urban crimes thus providing police with indications regarding new threats and patterns in criminal activity in the immediate aftermath of the pandemic. Similar reasoning could be applied to other global shocks that have great impact on daily routines of people (e.g., hurricane) and change the opportunity structure for offenders. Minimizing uncertainty in the face of emerging or persistent crime patterns will support effective an response to these challenges.

The paper develops around the following structure. Section 1 highlights the gap of knowledge addressed by this study and frames it within the criminological theoretical debate. Section 2 justifies the use of the COVID-19 emergency to answer our research question and outlines the methodology adopted and the data used for the analyses. Relying on two post-intervention time windows (the first one considering daily data points from March 4\textsuperscript{th} to March 16\textsuperscript{th}, the second one including all data points from March 4\textsuperscript{th} to March 28\textsuperscript{th}), the study dynamically compares the effects of milder containment policies prompted during the first two weeks of March with stricter measures put in place in the second half of the month. This design provides an informative framework in which to assess the evolution of the effects as a result of policy tightening in the immediate aftermath. Section 3 presents the statistical outcomes for the selected crime categories and the overall number of crimes in the city of Los Angeles. Finally, section 4 and section 5 review the most important results of the analyses, focusing on their theoretical interpretations and the most important policy implications entailed by the study.

\section{Research Background}
\label{background}

\subsection{Related Work}
Late modern history has been marked by the outbreak of several pandemics. In 1918, the so-called “Spanish Flu” infected about 27\% of the world’s population for an almost three-year period, with death estimates ranging from 17 to 50 million globally \citep{Taubenberger1918Influenzamother2006}. In 1957, the “Asian Flu” led to a total of 1.1 million deaths worldwide. In 1968, the A/H3N2 influenza shocked the world, causing about 1 million deaths in total. In 2002, the SARS emerged in China. During the period of infection, between November 2002 and July 2003, there were 8,098 reported cases of COVID-19 and 774 deaths. In 2009 the “Swine Flu” emerged from the United States and spread quickly around the world with an estimate of about 61 million cases in the United States alone. These public health emergencies modified and influenced many components of human society. Researchers investigated these changes from different standpoints. The interest of criminologists in pandemics is, instead, more recent as it mainly emerged in reaction to the spread of COVID-19.

Apart from a few studies focusing on the relation between SARS and suicides in Hong Kong \citep{ChanElderlysuicide20032006, Cheungrevisitolderadults2008}, research on crime and deviance is emerging only now in response to the COVID-19 pandemic. \cite{AshbyInitialevidencerelationship2020} used seasonal auto-regressive integrated moving average models to analyze trends in serious assaults (in public places and residences), burglaries (residential and non-residential), vehicle thefts in 16 large American cities. The analyses indicated no statistically significant changes in serious assaults. In some cities, \cite{AshbyInitialevidencerelationship2020} observed significant reductions in residential burglaries, but only minor changes in non-residential burglaries. Theft of motor vehicles also decreased in certain cities, while results are mixed when considering thefts from motor vehicles. \cite{MohlerImpactsocialdistancing2020} analyzed the counts of calls for service in Indianapolis and Los Angeles. The authors performed regressions in which they included an indicator for treatment–i.e., the period after the introduction of shelter in place orders–testing for differences in means of calls for six crimes in the period January 2 to March 16, 2020, which acted as baseline scenario. The results indicate a significant decrease in burglaries, robberies, and vandalism in Los Angeles and a significant increase in calls for service for domestic violence in both cities. Domestic violence is the focus also of the studies by \cite{PiqueroStayingHomeStaying2020} and by \cite{LeslieShelteringPlaceDomestic2020}. Piquero et al. (2020) identified a statistically significant increase in domestic violence in the first two weeks after the lockdown; yet, they also observed a subsequent decrease. Focusing on a sample of 15 metropolitan areas, \cite{LeslieShelteringPlaceDomestic2020} found that social distancing measures were associated with a 10\% increase in domestic violence service calls and observed that the increase might be actually higher due to underreporting.

Despite the importance of the topic, and the growing number of available studies, scientists have not yet answered many questions related to the possible effects of quarantine, social distancing, and self-isolation on crime and deviant behaviors. In particular, on the one hand, the short-term effect on crime of social-distancing policies characterized by different intensities on different types of crimes, from more instrumental ones–e.g., burglary–to more expressive and serious ones–e.g., homicide–is still to be determined. Few studies presented some results for Los Angeles (i.e., \cite{AshbyInitialevidencerelationship2020}; \cite{MohlerImpactsocialdistancing2020}), but they considered partially different crimes, partially different time frames, used different data (i.e., calls for service), and exploited different statistical methods (i.e., SARIMA model, regressions) compared to the analysis we are presenting here (i.e., nine categories of recorded crimes analyzed using Bayesian structural time-series models). The triangulation and comparison of different results are fundamental, given the novelty of the topic. At the same time, this analysis is the occasion to (indirectly) investigate the explanatory power of theories of crime as routine activity \citep{CohenSocialChangeCrime1979}, crime pattern theories \citep{BrantinghamPatternscrime1984} and general strain theory \citep{AgnewFoundationGeneralStrain1992} in the aftermath of such a major change in social interactions like the one due to the introduction of COVID-19 containment measures.

\subsection{Theoretical Framework and Hypotheses}
Routine activity \citep{CohenSocialChangeCrime1979} and crime pattern theories \citep{BrantinghamPatternscrime1984} stress that how the characteristics and interactions of individual-level activities command the spatial and temporal distribution of offending and victimization. On assuming these notions, members of a community can be modeled as potential offenders, potential victims, and potential guardians who move and interact in a socio-geographical space. Starting from these premises, routine activity theory postulates that offenders and victims–or targets–usually meet during everyday non-criminal activities \citep{BrantinghamPatternscrime1984}. Behavioral decisions then determine how the various agents react to each other’s presence and actions. Crime occurs in the context of the everyday routines as the three factors mentioned above converge in space and time: a motivated offender, a victim or potential target, and the absence of a capable guardian \citep{BrantinghamPatternscrime1984}.

Today, many criminological studies, especially those on ‘volume’ and urban crimes, rely on ideas emerging from theories that focus on situations and opportunities as triggers of crime \citep{CohenSocialChangeCrime1979, BrantinghamPatternscrime1984, BrantinghamCriminalityplace1995, ClarkeSituationalCrimePrevention1995, ClarkeSituationalCrimePrevention2009, WortleyEnvironmentalCriminologyCrime2008}. Because of the strong attention that these theories give to ordinary interaction in geographical and social space, we also rely on them to formulate our hypotheses on the short-term impact of the COVID-19-related social distancing measures. On the other hand, general strain theory postulates that stress generator factors like limited freedom of movement, strict physical and social isolation, in addition to economic uncertainty and concerns, may push youths, and people in general, to be more prone to commit crimes. These factors may introduce new negative stimuli while simultaneously removing positive ones thus generating negative feelings such as disappointment, depression, fear, and anger \citep{AgnewFoundationGeneralStrain1992}.

Overall, public measures intended to contain the spread of the virus cause people to spend more time at home and lose the density of social interactions. Accordingly, crime opportunities and places where crimes occur are likely to change from past observations and experiences. These changes can be particularly significant in the immediate aftermath of the health emergency. Under mild policies–i.e., from March 4\textsuperscript{th} to March 16\textsuperscript{th}–we expect a contraction in most urban crimes as the density of targets reduces in many areas of the city (Angel, 1968). We hypothesize crime reduction will be strong in crime generators areas, by which we mean “particular areas to which large numbers of people are attracted for reasons unrelated to any particular level of criminal motivation they might have or to any particular crime they might end up committing” \citep[p.~7]{BrantinghamCriminalityplace1995}. In Los Angeles, typical examples include the Hollywood entertainment district, the financial district with its high concentration of offices, and the famous Staple Center. Also flows of people to some crime attractor areas will be reduced because people have to avoid concentrating in bars, nightclubs, or shopping malls, but also in high-intensity drug trafficking and prostitution areas.

To various extents, we foresee a reduction in the number of batteries, assaults with deadly weapons, homicides, robberies, burglaries, shoplifting, thefts, and stolen vehicles as a consequence of a reduced interaction of people in the urban environment. Based on routine activity and crime pattern theories, we hypothesize that shoplifting diminishes the most. The reduction in the number of open shops and the limitations on the number of entrances–only a certain amount of people are allowed to be simultaneously in the shops depending on the premises’ square meters–reduce the opportunities for crime by simultaneously increasing the guardianship and reducing the exposure of targets to potential offenders.

Routine activity \citep{CohenSocialChangeCrime1979} and crime pattern theories \citep{BrantinghamPatternscrime1984} suggest that stricter social distancing policies–i.e., the period from March 16\textsuperscript{th} to March 28\textsuperscript{th}–should have a stronger impact on the crimes considered than mild policies. This is due to the further reduction in movement and social interaction induced by the reinforcement of social distancing measures. At the same time, the prolonged stay-at-home order may trigger increases in types of crimes in the medium- and eventually long-term as effects of the increased stress to which people are exposed. While according to the general strain theory all types of crime may be influenced by this dynamic \citep{AgnewBuildingFoundationGeneral2001a}, in line with the findings of \cite{SchoepferSelfControlMoralBeliefs2006}, expressive crimes are more likely to be directly influenced by the increased strain that people experience.
Finally, differently from the other crimes considered, intimate partner assaults are likely to increase. Patriarchy and gender inequality are often considered to be the root causes of intimate partner violence; yet, situational determinants are also recognized as influencing this form of crime \citep{WilkinsonSituationaldeterminantsintimate2005}. As a consequence of the spread of COVID-19, couples–including dysfunctional ones–spend more time together in their homes with a reduced presence of possible informal guardians like relatives and acquaintances, two factors that may lead to an increase in violence  outbreaks (Hayes, 2018). At the same time, the strain caused by the pandemic makes people more likely to respond with anger to confrontations and to be less concerned about hurting others thus possibly boosting violent crimes \citep{AgnewBuildingFoundationGeneral2001a, BroidyTestGeneralStrain2001}, both in the period under mild policies and even more in that under stricter ones.

\section{Analytical Framework}
\label{analytical}

\subsection{Methodology}
Evaluating the causal link and impact of certain policies is a crucial aspect of research and practice. Criminologists working on different topics have long attempted to assess the extent to which public interventions aimed at reducing crime are actually effective in fulfilling their mission. The standard for assessing the causal impact of a certain intervention is represented by Randomized Controlled Trials (RCT) \citep{RubinEstimatingcausaleffects1974}. However, this research design is often unfeasible due to issues related to financial costs, ethics or practical obstacles–e.g., complex regulatory requirements.
While \cite{PearlCausalinferencestatistics2009} demonstrates that post-facto observational studies cannot provide evidence of causal inference due to the potential presence of confounding factors, several quasi-experimental alternatives have been proposed to overcome the difficulty of running RCT in certain scientific fields. Among these methods are interrupted time series. Interrupted time series have gained popularity in sociology and criminology, and they have been applied to several different research problems \citep{BiglanValueInterruptedTimeSeries2000, HumphreysEvaluatingImpactFlexible2013, PridemoreReductionMaleSuicide2013, PridemoreEffects2006Russian2014, HumphreysEvaluatingImpactFlorida2017}. This method makes it possible to assess the effect of a certain policy by analyzing the change in the level and slope of the time series after an intervention has been applied, compared to the structure of the temporal dynamic before the intervention. More recently, scientists have developed a framework for evaluating the causal influence of a certain intervention relying on Bayesian statistics.
Following this later evolution, this work investigates the effect of social distancing and related measures in the attempt to contain COVID-19 on criminal trends in Los Angeles using Bayesian structural time-series (BSTS) models \citep{BrodersenInferringcausalimpact2015}. Specifically, we apply a method relying on diffusion regression state-space which predicts a counterfactual trend in a synthetic control that would have occurred in a virtual counterfactual scenario with no intervention–thus, in a scenario where no containment policies are promulgated. This approach allows us to quantify the short-term impact and statistical significance of the containment policies on our variable of interest, namely the number of crimes over time. BSTS are state-space models specifically defined by two equations. The first, i.e. the observation equation, being:

\begin{equation}
    y_{t}=Z_{t}^{\mathrm{T}}\alpha_{t} +\varepsilon _{t}
\end{equation}
where $y_t$ is a scalar observation, $Z_t$ is the $d$-dimensional output vector and $\varepsilon_t \sim \mathcal{N} (0, \sigma_{t}^{2} )$ and $\varepsilon _{t}$ is a scalar observation error with noise variance $\sigma_t$. The \textit{observation equation} connects the observed data $y_t$ to a latent $d$-dimensional state vector $\alpha_{t}$. The second, equation, instead is the \textit{state equation}, which reads:
\begin{equation}
    \alpha _{t+1}=T_{t}\alpha _{t}+R_{t}\eta _{t}
\end{equation}
where $T_t$ is a $d \times d$ transition matrix, $R$ is a $d \times q$ control matrix, $\eta _{t}$ is a $q$-dimensional systems error with a $q \times q$ state diffusion matrix $Q_t$ such that $\eta_t \sim \mathcal{N}_t (0, Q_{t})$.
This second equation specifically governs the dynamic change of the state vector $\alpha _{t}$ through time. The inferential dimension in the model comprises three components. First, draws of the model parameters $\theta$ and the state vector $\alpha$ (given $\mathbf{y}_{1:n}$, i.e. the observed data in the training period) are simulated. Second, the model uses posterior simulations to simulate from $p(\tilde{\mathbf{y}}_{n+1:m}\mid \mathbf{y}_{1:n})$, which is the posterior predictive distribution, with $\tilde{\mathbf{y}}_{n+1:m}$ as the counterfactual time series and $\mathbf{y}_{1:n}$ as the observed time series before the intervention. Third, using posterior predictive samples, the model compute the posterior distribution of the point-wise impact $y_t-\tilde{y}_t$ for each time unit $t$.

The Bayesian framework in which the model is embedded allows flexibility and inferential power, enabling the method to effectively estimate the cumulative difference between the actual data and a counterfactual scenario. The proposed modeling architecture, through the comparison between a univariate and a multivariate model (which includes two covariates, i.e., daily temperature and presence of holidays) and the exploitation of a long timeframe, controls the risk of excluding relevant patterns that may not be specifically related to the pandemic and avoid the risk of ignoring long-term dynamics, a pitfall that would lead to biased estimates. Moreover, the weekly component embedded in the estimation technique preserves the inherent seasonal component often exhibited by criminal activity.

Concerning the implementation part, we have relied on the CausalImpact package available in R. Each model has been fit by only considering a single target time-series mapping the trend of crime categories in the time window under consideration at a time. The package does not allow to simultaneously model multiple target variables and, having included a crime series as the target one and all the others as covariates would have contradicted the requirement to only consider covariates that are not influenced by the interventions.  Each model included a seasonal component: characterizing weekly seasonality (number of seasons set equal to 7 with season duration equal to 1, i.e., one day). This allowed to account for the well-known seasonal oscillations of crime over the days of a week (with temperature as a covariate being able to control higher-level seasonal components). , We performed our analyses by including weekly seasonal terms to account for weekly variations in crime trends. By doing so, it was possible to consider also possible seasonal variations that do not directly depend on temperature-related mechanisms, but instead relate to routine activities of people and places during social occurrences (e.g. school closure) \citep{AndresenCrimeseasonalityits2013}. Furthermore, to obtain accurate estimates for each model a total of 1,000 Markov Chain Monte Carlo (MCMC) samples are drawn. Finally, the prior of each model, expressed in terms standard deviation of the Gaussian random walk at the local level, has been kept equal to 0.1 as suggested by the authors of the package in absence of ground truth. The value represents a good compromise between a high standard deviation that would assume that the variations in the signal are all explained by the intervention and a very small standard deviation that imposes that such variations are instead solely due to high noise in the data.\footnote{The R code used to perform the analyses here presented is publicly available in a dedicated repository at \href{https://github.com/gcampede/Covid-19_and_crime_LA}{Link}.}

\subsection{Time Frame}
\begin{figure}[hbt!]
    \centering
    \includegraphics[scale=0.42]{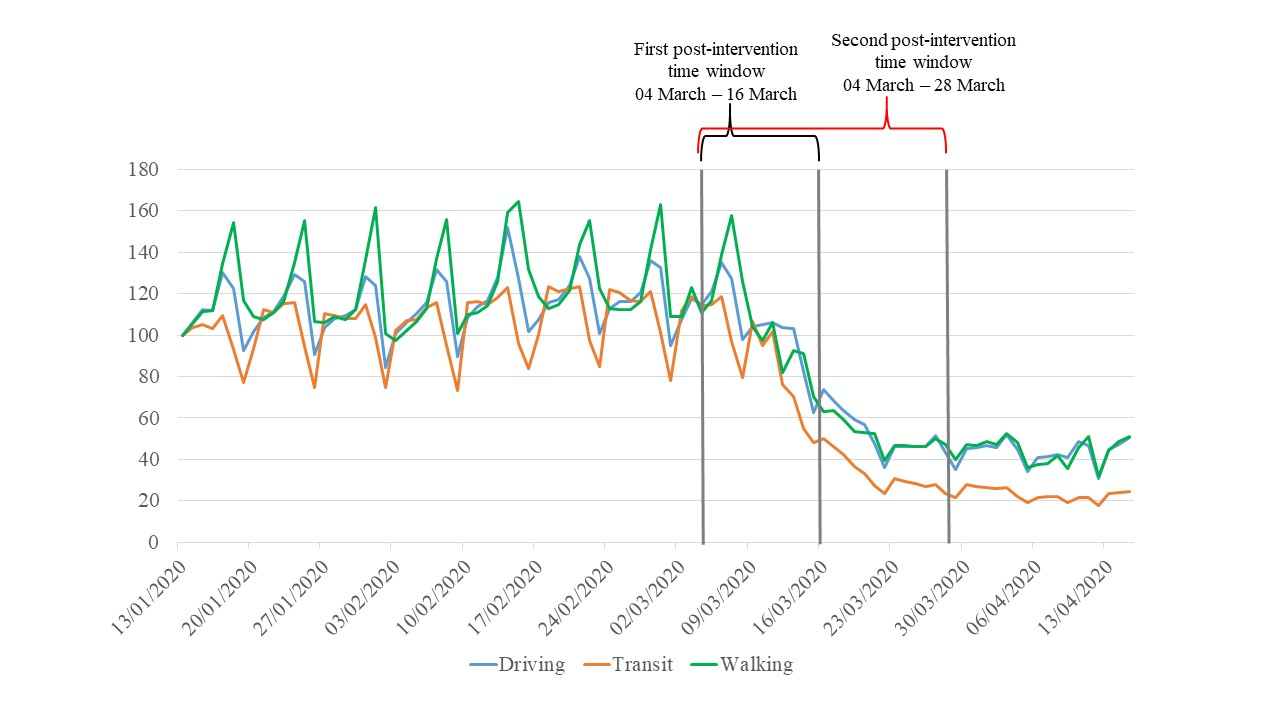}
    \caption{Mobility trends in the city of Los Angeles, January 1\textsuperscript{st} 2020 – April 15th 2020. Source: Mobility Trends Reports, Apple (2020)}
    \label{fig:mobility}
\end{figure}

Exploiting this analytical strategy, we separately analyze the period up to March 16\textsuperscript{th} to test the effect of milder containment policies (the first post-intervention time window goes from March 4\textsuperscript{th} to March 16\textsuperscript{th}) and up to March 28\textsuperscript{th} to test the effect of the introduction of stricter containment policies (the second post-intervention time window goes from March 4\textsuperscript{th} to March 28\textsuperscript{th}).\footnote{The two post-intervention time windows considered are not mutually exclusive. The first one includes the days between the 4\textsuperscript{th} of March to the 16\textsuperscript{th} of March and the second one considers the days between the 4\textsuperscript{th} of March to the 28\textsuperscript{th} of March. The second time window includes the first one. Our interest is to map the evolution of the policies and their influence on crime trends given that policy effects in this context are cumulative rather than mutually exclusive. The focus of our analysis is not to compare two different time period of post intervention in order to understand, for example, if mild policies are better than harsh policies, but to assess the effect of those policies on crime over time.}  These two post-intervention time-windows enable us to obtain a more comprehensive and inherently dynamic description of the reality. In fact, given the progressive tightening of imposed restrictions, avoiding a comparative analysis of the evolution of the effects over different weeks would somewhat oversimplify and overly aggregate our statistical estimates. Conversely, the possibility to observe the evolution of the effects helps in assessing the strength of a trend, anticipating the likely developments and better explaining its behavior  by framing it in our theories of reference. Figure \ref{fig:mobility} presents the data on people’s mobility (driving, transit and walking) in the city of Los Angeles from January 1\textsuperscript{st} 2020 to April 15th 2020 and shows how in the first post-intervention time window the mobility has already started to reduce compared to the previous months. This trend definitively continued after March 16\textsuperscript{th} when stricter policies were implemented.\footnote{The mobility trends in Los Angeles are also confirmed by other studies that used other data to analyze trends in commuting and general mobility (e.g., data provided by Cuebiq) \citep{KleinAssessingchangescommuting2020, Ruiz-EulerMobilityPatternsIncome2020}.}  Our interest is to understand changes in the crime patterns in the immediate aftermath of the health emergency to assess and analyze shot-term dynamics. Indeed, in the medium- and long-run other factors might play a more central role in explaining the changes in crime trends (e.g., unemployment, social riots).

\subsection{Data}
To conduct our analysis, we first drew upon the (\href{https://data.lacity.org}{Los Angeles Open Data} portal). Unlike prior studies that analyzed the counts of calls for service in Los Angeles (e.g., \cite{MohlerImpactsocialdistancing2020}), we used data on crime reported in the city. Exploiting the website API, we accessed two different datasets on crimes reported by the Los Angeles Police Department (LAPD). The first one comprised all crime incidents reported from 2017 to 2019. The second one referred to all the crimes reported from January 1\textsuperscript{st} to March 30th 2020. The two datasets contain detailed daily observations regarding each reported crime with information on the type of offense–organized in 140 crime categories–the age, gender, and descendants of the victim (if any), the type of weapon (if any), and the location of the occurrence.
\newpage
\begin{table}[!hbt]
\centering
\footnotesize
\begin{tabular}{lcc}
\hline
\textbf{Dataset} & \multicolumn{1}{l}{\textbf{First Day Considered}} & \textbf{N of Observations} \\\hline\hline
2017-2019 & 2017-01-01 & 685,615 \\\hline
2020-onwards & 2020-01-01 & \begin{tabular}[c]{@{}c@{}}47,252 \\ (updated at 2020-03-30, \\ includes data up to 2020-03-28)\end{tabular} \\\hline
Merged & 2017-01-01 & \begin{tabular}[c]{@{}c@{}}732,867\\ (updated at 2020-03-30 \\ includes data up to 2020-03-28)\end{tabular}\\\hline
\end{tabular}
\caption{Number of Observations and Starting Point per Dataset}
\label{desc}
\end{table}

\begin{figure}[!hbt]
    \centering
    \includegraphics[scale=0.50]{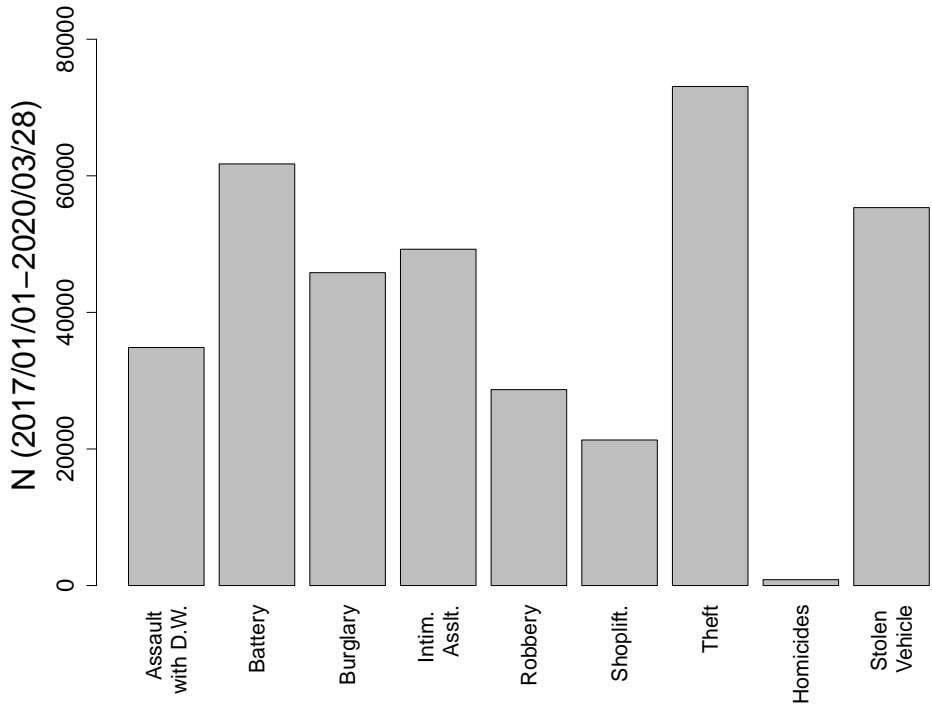}
    \caption{Number of Observations per Crime Category}
    \label{fig:category}
\end{figure}

For the purpose of the present work, we only relied on the reported date of crime occurrence and on the offense categories as meaningful sources of information to analyze city-wide criminal trends. A brief description of the three datasets (1. the one comprising crimes from 2017 to 2019; 2. the one with offenses up to March 2020; 3. the merged one used in the models) is provided in Table \ref{desc}. We extracted the crime categories of interest and we then grouped observations by daily counts, obtaining separated time series for each crime category. While our models focus on the post-intervention period from March 4\textsuperscript{th} to March 28\textsuperscript{th}, we lastly accessed data on April 7th, thus ensuring that a larger number of crime reports that were not possibly included in the dataset in their first week after occurrence were actually imputed by the LAPD soon thereafter. Figure \ref{fig:category} displays the number of observations per crime category in the period from January 1\textsuperscript{st} 2017 to March 28\textsuperscript{th} 2020. Thefts are the most frequent crimes, followed by battery and stolen vehicles. Homicide is by far the least prevalent offense in the sample but, trivially, the most serious one.

Furthermore, to get controls we have used two additional datasets. The first one is an open-access dataset provided by the governmental site of the National Centers for Environmental Information (\href{https://www.ncdc.noaa.gov/}{https://www.ncdc.noaa.gov}) with data on minimum, average and maximum temperature daily registered at the Los Angeles Airport in the timeframe of interest. The second one is a publicly available dataset containing American holidays. 

The literature has long found relationships between temperature and variation in crime trends \citep{CohnWeatherCrime1990, FieldEffectTemperatureCrime1992} and between festivities and holidays and criminal temporal clustering \citep{CohnEvencriminalstake2003, TowersFactorsinfluencingtemporal2018}. It is worth specifying that we do not seek to establish any connection between the theories that we exploit to frame the present research problem and the selected covariates. These covariates have to be interpreted as diagnostic measures aimed at testing the reliability of the statistical estimates obtained in the univariate models. Temperature and holidays play a role in smoothing and controlling the potential presence of noise in the signals of criminal trends (especially given the relevant duration of the time frame under consideration). We hence construct alternative models with covariates to ensure that the estimates obtained with no controls are not biased due to the exclusion of relevant confounders.

As required by the method, these two covariates have not been affected by the policies under scrutiny. This is trivial in the case of average temperature, while holidays may be affected in the future if restrictions are prolonged for months. Yet, in the period under analysis, dates of holidays were not changed because of the pandemic. The method adopted generally recommends using more than two controls to evaluate the effect of an intervention on the response time series. However, we can assume that no other daily predictor of crime has been strongly influenced by the containment policies during the period analyzed. 
Differently from other studies on the topic, the use of time series spanning 39 months was made to reduce the potential biases arising from the exclusion of hidden trend dynamics, preserving the seasonality and long-term dependencies of crime.  Table \ref{ts_desc} reports the main descriptive statistics for the time series that are part of the analysis.

\begin{table}[!hbt]
\centering
\footnotesize
\begin{tabular}{lccccccc}
\hline
\textbf{Variable} & \textbf{Min} & \textbf{1\textsuperscript{st} Q} & \textbf{Median} & \textbf{Mean} & \multicolumn{1}{l}{\textbf{St. Dev.}} & \textbf{3rd Q} & \textbf{Max} \\ \hline\hline
All Crimes & 254.00 & 570.00 & 628.00 & 619.00 & 69.80 & 669.00 & 796.00 \\ \hline
Assault (with Deadly Weapon) & 6.00 & 24.00 & 29.00 & 29.44 & 7.53 & 34.00 & 61.00 \\ \hline
Battery & 15.00 & 46.00 & 52.00 & 52.15 & 9.51 & 58.00 & 93.00 \\ \hline
Burglary & 13.00 & 31.00 & 38.00 & 38.70 & 9.96 & 45.00 & 93.00 \\ \hline
Homicide & 0.00 & 0.00 & 0.00 & 0.72 & 0.00 & 1.00 & 5.00 \\ \hline
Intimate Partner Assault & 10.00 & 36.00 & 41.00 & 41.60 & 8.84 & 47.00 & 78.00 \\ \hline
Robbery & 7.00 & 20.00 & 24.00 & 24.23 & 5.85 & 28.00 & 48.00 \\ \hline
Shoplifting & 2.00 & 15.00 & 18.00 & 18.00 & 4.92 & 21.00 & 33.00 \\ \hline
Theft & 19.00 & 52.00 & 62.50 & 61.72 & 13.3 & 61.72 & 71.00 \\ \hline
Stolen Vehicle & 19.00 & 40.00 & 46.50 & 46.74 & 9.39 & 52.00 & 88.00 \\ \hline
Holiday & 0.00 & 0.00 & 0.00 & 0.02 & 0.16 & 0.00 & 1.00 \\ \hline
Max Temperature & 52.00 & 69.00 & 76.00 & 75.84 & 9.20 & 82.00 & 108.00 \\ \hline
\end{tabular}
\caption{Descriptive Statistics of the Considered Time Series}
\label{ts_desc}
\end{table}

\section{Results} \label{results}
\begin{figure}[!hbt]
    \centering
    \includegraphics[scale=0.53]{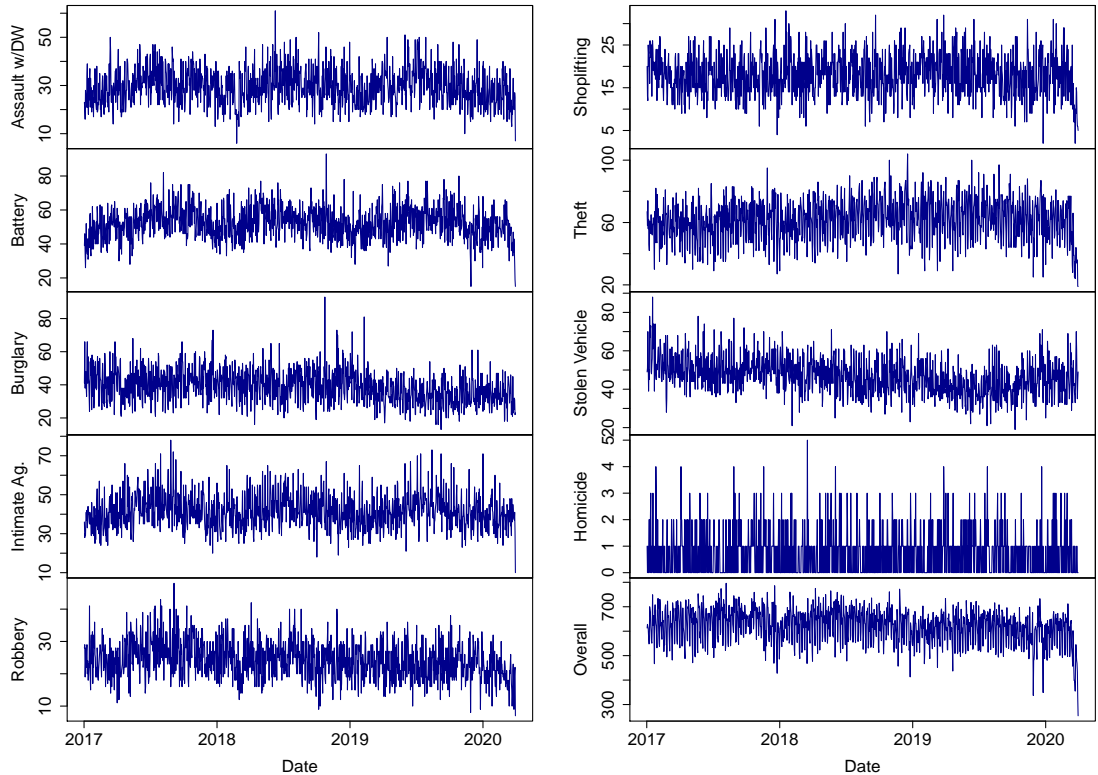}
    \caption{Time Series of Considered Crimes}
    \label{fig:series}
\end{figure}

This section presents the results on the causal impact of the policy interventions starting on March 4\textsuperscript{th} per each offense type and for the overall number of crimes. For each crime, the same analytical structure is provided. We ran two different models. The first one was a univariate model that only considers the time series of interest without controls. The second model integrates two covariates to control for spurious effects and unobservable dynamics. At the same time, the two models were performed on two time-windows; the first capturing the impact of mild policies (from March 4\textsuperscript{th} to March 16\textsuperscript{th}), the second including also the first weeks in which stricter policies entered into force (from March 4\textsuperscript{th} to March 28\textsuperscript{th}). Figure \ref{fig:series} shows the time series of the crimes considered; while, to facilitate the reading of the results and summarize the statistical outcomes of the models, Table \ref{generalresults} presents all the statistical results (full statistical outcomes are available in the Supplementary Materials, from Table \ref{assaults} to \ref{all}).

\begin{table}[!hbt]
\footnotesize
\centering
\begin{tabular}{l|cc|cc}
\hline
\multirow{2}{*}{\textbf{Crime Type}} & \multicolumn{2}{c|}{\textbf{\begin{tabular}[c|]{@{}c@{}}First post intervention \\ time window\\ (March 4\textsuperscript{th} – March 16\textsuperscript{th})\end{tabular}}} & \multicolumn{2}{c}{\textbf{\begin{tabular}[c]{@{}c@{}}Second post intervention \\ time window \\ (March 4\textsuperscript{th} – March 28\textsuperscript{th})\end{tabular}}} \\ \cline{2-5} 
 & \textit{Univariate} & \textit{With Covariates} & \textit{Univariate} & \textit{With Covariates} \\ \hline\hline
Assaults D.W. & \begin{tabular}[c]{@{}c@{}}-2.98\%   \\ {[}-19\%, 13\%{]}\end{tabular} & \begin{tabular}[c]{@{}c@{}}-1.5\%   \\ {[}-18\%, 13\%{]}\end{tabular} & \begin{tabular}[c]{@{}c@{}}-11\%**   \\ {[}-23\%, 2.8\%{]}\end{tabular} & \begin{tabular}[c]{@{}c@{}}-6.3\% (6\%)   \\ {[}-18\%, 5.5\%{]}\end{tabular} \\ \hline
Battery (Simple Assault) & \begin{tabular}[c]{@{}c@{}}-0.6\%   \\ {[}-12\%, 11\%{]}\end{tabular} & \begin{tabular}[c]{@{}c@{}}0.78\% \\ {[}-9.2\%, 11\%{]}\end{tabular} & \begin{tabular}[c]{@{}c@{}}-11\%**  \\ {[}-21\%, -0.99\%{]}\end{tabular} & \begin{tabular}[c]{@{}c@{}}-7.6\%**   \\ {[}-16\%, 0.39\%{]}\end{tabular} \\ \hline
Burglary & \begin{tabular}[c]{@{}c@{}}0.89\%  \\ {[}-14\%, 15\%{]}\end{tabular} & \begin{tabular}[c]{@{}c@{}}-0.58\%  \\ {[}-14\%, 11\%{]}\end{tabular} & \begin{tabular}[c]{@{}c@{}}-4.8\% \\ {[}-15\%, 5.5\%{]}\end{tabular} & \begin{tabular}[c]{@{}c@{}}-7.3\%*  \\ {[}-17\%, 3.3\%{]}\end{tabular} \\ \hline
Intimate Partner Assault & \begin{tabular}[c]{@{}c@{}}-4\%   \\ {[}-16\%, 6.4\%{]}\end{tabular} & \begin{tabular}[c]{@{}c@{}}-2.5\% \\ {[}-13\%, 8.6\%{]}\end{tabular} & \begin{tabular}[c]{@{}c@{}}-0.28\% \\ {[}-11\%, -11\%{]}\end{tabular} & \begin{tabular}[c]{@{}c@{}}3.3\%  \\ {[}-5.6\%, -12\%{]}\end{tabular} \\ \hline
Robbery & \begin{tabular}[c]{@{}c@{}}-24\%***   \\ {[}-38\%, -8.5\%{]}\end{tabular} & \begin{tabular}[c]{@{}c@{}}-23\%***   \\ {[}-38\%, -8.7\%{]}\end{tabular} & \begin{tabular}[c]{@{}c@{}}-21\%***  \\ {[}-33\%, -9.3\%{]}\end{tabular} & \begin{tabular}[c]{@{}c@{}}-19\%***  \\ {[}-30\%, -8.7\%{]}\end{tabular} \\ \hline
Shoplifting & \begin{tabular}[c]{@{}c@{}}-14\%*** \\ {[}-30\%, 2.4\%{]}\end{tabular} & \begin{tabular}[c]{@{}c@{}}-15\%***\\ {[}-30\%, 0.34\%{]}\end{tabular} & \begin{tabular}[c]{@{}c@{}}-31\%***  \\ {[}-42\%, -20\%{]}\end{tabular} & \begin{tabular}[c]{@{}c@{}}-32\%****   \\ {[}-43\%, -21\%{]}\end{tabular} \\ \hline
Theft & \begin{tabular}[c]{@{}c@{}}-9.1\%**\\ {[}-19\%, 0.57\%{]}\end{tabular} & \begin{tabular}[c]{@{}c@{}}-9.6\%**\\ {[}-19\%, -1\%{]}\end{tabular} & \begin{tabular}[c]{@{}c@{}}-24\%***  \\ {[}-31\%, -17\%{]}\end{tabular} & \begin{tabular}[c]{@{}c@{}}-25\%***   \\ {[}-31\%, -18\%{]}\end{tabular} \\ \hline
Stolen Vehicles & \begin{tabular}[c]{@{}c@{}}1\%\\ {[}-9.4\%, 11\%{]}\end{tabular} & \begin{tabular}[c]{@{}c@{}}0.06\%   \\ {[}-10\%, 9.9\%{]}\end{tabular} & \begin{tabular}[c]{@{}c@{}}1.5\%   \\ {[}-6.5\%, 9.6\%{]}\end{tabular} & \begin{tabular}[c]{@{}c@{}}-0.12\%\\ {[}-7.4\%, 7.5\%{]}\end{tabular} \\ \hline
Homicides & \begin{tabular}[c]{@{}c@{}}-15\% \\ {[}-88\%, 57\%{]}\end{tabular} & \begin{tabular}[c]{@{}c@{}}-10\%   \\ {[}-84\%, 59\%{]}\end{tabular} & \begin{tabular}[c]{@{}c@{}}-28\%   \\ {[}-79\%, 25\%{]}\end{tabular} & \begin{tabular}[c]{@{}c@{}}-24\%  \\ {[}-76\%, 31\%{]}\end{tabular} \\ \hline
Overall Crimes & \begin{tabular}[c]{@{}c@{}}-5.6\%*** \\ {[}-10\%, -1.5\%{]}\end{tabular} & \begin{tabular}[c]{@{}c@{}}-5.4\%**  \\ {[}-9.5\%, -1\%{]}\end{tabular} & \begin{tabular}[c]{@{}c@{}}-15\%***   \\ {[}-18\%, -11\%{]}\end{tabular} & \begin{tabular}[c]{@{}c@{}}-14\%***   \\ {[}-17\%, -11\%{]}\end{tabular} \\ \hline
\end{tabular}
\caption{Model Results - Relative Cumulative Effect per Each Crime (95\% C.I. Between Parentheses)}
\label{generalresults}
\end{table}

\subsubsection*{Assault with Deadly Weapons}
Concerning assaults with deadly weapons,\footnote{\textit{Assault with Deadly Weapons} corresponds to crimes labelled as Assault with Deadly Weapon, Aggravated Assault by the LAPD. According to the California Penal Code 245(a)(1), an assault with a deadly weapon occurs when an individual wrongfully attacks a victim with an object that can seriously injure or even inflict death.}  both the univariate and the multivariate models, for both time windows, i.e. up to March 16\textsuperscript{th} and up to March 28\textsuperscript{th}, do not show any statistically significant effect of the containment policies. Although all the models report negative coefficients, indicating a reduction in the trend in absolute and relative terms, the posterior probability of a causal effect is, respectively, 64\% and 52\% for the first time window and 94\% and 85\% for the second time window. The results provided by the models referring to the second time window–especially for the univariate one–are much closer to a statistically significant outcome, suggesting a different effect when stringent policies are introduced. Consequently, it is possible that this negative effect will show statistically significant results in the case of prolonged or even more stringent containment measures. 

Table \ref{assaults} reports the outcomes of the analysis.

\subsubsection*{Battery and Simple Assault}

The occurrence of batteries\footnote{Although \textit{Battery and Simple Assault} are coded as a single type of offense one in the original database under the crime category BATTERY - SIMPLE ASSAULT, California assault law, disciplined by Penal Code 240PC, provides two distinct definitions for battery and assault. A simple assault is the attempt to use force or violence against someone else, while battery is the actual use of force or violence against one or more individuals.} is affected by strict policies, while it is not so by loose ones. Up to March 16\textsuperscript{th}, containment policies do not seem to have resulted in a reduction of battery and simple assaults. Conversely, if we consider the introduction of more stringent measures of social distancing, it is possible to observe a statistically significant reduction of this crime. The models considering crimes recorded until March 28\textsuperscript{th} show a relative effect of -11.0\% in the univariate model and of -7.6\% in the multivariate one. The posterior probability of a causal effect is, respectively, 98\% and 96\% indicating statistically significant outcomes for both models (full results are available in Table \ref{battery}).

\subsubsection*{Burglaries}

For what concerns burglaries,\footnote{In California, according to the Penal Code, burglary is defined as the act of entering any structure, room, or locked vehicle with the intention to commit a theft or a felony. Furthermore, an individual can be considered guilty of burglary even if the intended crime, once entered, is never been committed. Our initial hypothesis is that containment policies should have a clear impact on this crime category as people encouraged or forced to stay in their houses increase guardianship, thus reducing the chance for an individual to enter a property without being noticed.}  the policies to contain the spread of COVID-19 have not produced any significant effect in the first four weeks from their introduction. Although we expected a slight decrease in their overall occurrence, given the extensive agreement over the fact that burglars prefer to target unoccupied homes \citep{ShoverBurglary1991,MustaineVictimizationrisksroutine1997, TseloniModellingPropertyCrime2002} and the consequent increased guardianship enforced by people staying at home due to the pandemic, statistical outcomes do not corroborate our hypothesis. In the first time window, the effects are minimal and the results of the two models diverge in their directions. The univariate model shows an increase of 0.89\% with respect to the predicted value in the absence of an intervention, while the multivariate displays an effect equal to -0.58\%, with posterior probabilities respectively being 56\% and 50\%. The models performed up to March 28\textsuperscript{th} show a non-significant reduction of burglaries in the city with a higher posterior probability for both the univariate and the multivariate models–respectively, 80\% and 91\%–compared to the other time-frame analyzed. Full results can be found in Table \ref{burglary}.

\subsubsection*{Intimate Partner Assault}

COVID-19-containment policies, we hypothesize, could cause intimate partner assaults\footnote{To analyze intimate partner assaults, we have combined both simple and aggravated assaults–Intimate Partner - Aggravated Assault” and “Intimate Partner - Simple Assault--- in the original database by LAPD. These two offenses fall within the broader set of domestic violence crimes, which includes other forms of within-family violence–e.g., parents being violent against their children. The California Penal Code defines an intimate partner as a current or former spouse, a fiancé, a co-parent of a child, a person with whom the perpetrator had a dating relationship or a person with whom the perpetrator lives.}  to increase as a consequence of individuals spending longer time at home in a potentially stressful situation. Instead, results suggest that the policies adopted have not prompted any immediate significant change in intimate partner assaults. The models considering the days from March 4\textsuperscript{th} to March 16\textsuperscript{th} as period of intervention show non-significant negative effects (-4.0\% and -2.5\%). The univariate model considering the entire period identifies a small non-significant negative effect (-0.28\%). Finally, the multivariate model indicates an increase in intimate assaults due to the policies (+3.3\%); but this increase is not statistically significant. The explanation for the absence of a clear and significant signal in the post-intervention period may be connected to the fact that dynamics of this crime are complex and the forced cohabitation of partners is not an immediate trigger of violence within the household. While in the case of shoplifting, for example, the closure of shops has a direct impact on the thefts within the shops themselves, in the case of intimate partner violence the increase may be delayed as tensions intensify.

\subsubsection*{Robbery}

Robbery\footnote{Section 211 of the California Penal Code defines robbery as the act of taking personal property from someone else, against the targeted victim’s will, using force or fear. Robberies are classified as felonies. Data on robberies correspond to \textit{Robbery} in the original database compiled by the LAPD.}  shows a significant change in the post-intervention period. This applies to both the univariate and the multivariate cases for both the temporal windows selected. In the univariate case, the relative effect is estimated as a reduction of 24\% in robberies, with a cumulative total of 202 criminal events against a predicted 266 in a non-intervention scenario in the first time window. In the second time window, the estimated effect is a reduction of 21\%, with a cumulative total of 439 robberies against a predicted 533 in a non-intervention scenario (See Table \ref{robbery} in the Supplementary Materials for the Absolute Effect). In the multivariate case, the effects are slightly reduced in both the time frames; being respectively, -23\% and -19\%. The magnitude of the effects is thus high. The posterior probability is 99.7\% and 99.8\%, respectively, for the univariate and multivariate models up to March 16\textsuperscript{th} and 99.7\% and 99.8\% for the models up to March 28\textsuperscript{th}.

\subsubsection*{Shoplifting}

The statistical outcomes of the models referring to shoplifting\footnote{According to the California Penal Code 459.5, shoplifting is the offense of entering a commercial establishment during business regular hours, with the intent of committing a theft crime worth \$950 or less, regardless of the actual completion of the theft. This crime is identified as \textit{Shoplifting-Petty Theft (\$950 \& Under)}  in the LAPD’s database. In the present work, we have also considered shoplifting grand-thefts, related to attempted thefts of property worth more than 950\$–i.e., \textit{Shoplifting-Grand Theft (\$950.01 \& Over)}  in the original dataset.}  indicate a significant reduction after the introduction of the state of the emergency in Los Angeles. The results hold for all models. In the first time window analyzed, 189 shoplifting cases were registered, significantly fewer than the cumulative numbers of occurrences predicted by the virtual scenarios with no intervention: 220 for the univariate and 223 in the multivariate cases (Table \ref{shoplifting} in the Supplementary Materials). In the models up to March 28\textsuperscript{th}, against cumulative predicted values of 462 and 471 in a scenario without intervention, 320 shoplifts were recorded by the LAPD. This indicates an estimated relative reduction of 14\% in the univariate case, and a 15\% reduction in the multivariate one in the period with mild policies. With the introduction of stricter policies, the relative reduction is 31\% in the univariate case, and 32\% in the multivariate one. The probability of the effects being a causal consequence of the policies is 95.6\% and 97.2\% for the first time window and 99.7\% and 99.9\% for the second time window, providing strong statistical evidence.

\subsubsection*{Theft}
In line with what we found for robbery and shoplifting, thefts\footnote{For the purpose of this study, we have combined in the general category “Theft” both petty thefts–i.e., \textit{Theft Plain-Petty (\$950 \& Under)} –and grand thefts–i.e., \textit{Theft-Grand (\$950.01 \& Over) Excpt, Guns, Fowl, Livestk, Prod}. The California Penal Code defines petty theft as the act of stealing–or wrongfully taking–an object belonging to someone else when the value of the property is equal to \$950 or less. Grand theft, instead, is a more serious offense and pertains to acts in which the property has a value higher than 950\$.}  also appear to be affected by the application of the COVID-19-related containment measures. All models show a statistically significant reduction of thefts. In the first analyzed window, the models report a 9.1\% decrease in the univariate case (significant at the 96.7\% level) and a 9.6\% reduction in the multivariate one (significant at the 98.3\%). In the second window considered (up to March 28\textsuperscript{th}), the models estimate a 21\% decrease in the univariate case (significant at the 99.7\%) and a 25\% reduction in the multivariate case (significant at the 99.9\%).

\subsubsection*{Stolen Vehicles}
The policies have not had any significant effect on vehicle thefts.\footnote{\textit{Stolen vehicles} refers to two different offenses, namely grand theft auto and the unlawful taking or driving of a vehicle. The main difference between the two offenses pertains to the duration of the crime itself. If, for instance, a person steals a car with the intent to keep it, this is often considered grand theft auto. Conversely, if the offender aims at using the car for a ride–or, in any case, for a short timeframe–the act is usually considered as the unlawful taking or driving of a vehicle. The original database compiled by the LAPD provides a single crime category: \textit{Vehicle - Stolen}.}  Indeed, the relative effects are not only not significant, but also small for all models: 1.0\% and 1.5\% in the univariate cases, 0.06\% and -0.12\% in the multivariate ones. There are different possible (complementary) explanations for this finding. On the one hand, while at home, a car owner may be an ineffective guardian of her/his own car. On the other hand, because the theft of cars is often related to their immediate use \citep{CherbonneauAutoTheft2011}, the slowdown of productive and social activities–including other crimes–may mean that offenders have less need to steal a vehicle. The combination of these dynamics, which are pushing in opposite directions, might explain the absence of a clear impact of social distancing policies on stolen vehicles in Los Angeles.

\subsubsection*{Homicides}
Overall, COVID-19 containment policies do not show any statistically significant effect on homicides\footnote{The category\textit{Homicides} comprises those crimes which are disciplined by California’s Homicide Laws and corresponds to the Criminal Homicide crime category in the original database. A person committing a homicide can be prosecuted in several ways depending on the characteristics of the action. Among these are first-degree murder, second-degree murder, capital murder, voluntary manslaughter, involuntary manslaughter and vehicular manslaughter. Notably, homicides are the most serious offenses among those considered in this study–and the least common.}  in the short aftermath of their deployment. In neither of the two selected temporal windows do the models detect a sufficient statistically strong variation in the trend. Nonetheless, compared to the first batch of data–up to March 16\textsuperscript{th}–the models also considering the strict policies point in a two-fold direction. First, the relative effect has increased, ranging from -15\% to -28\% according to the univariate model and from -10\% to -24\% according to the multivariate specification. Second, the posterior tail-area probability p has decreased, thus increasing the posterior probability of a causal effect, which ranged from 67\% and 63\% in the first batch of results, to 86\% and 80\%.

\subsubsection*{Overall Crimes}
\begin{figure}[!hbt]
    \centering
    \includegraphics[scale=0.55]{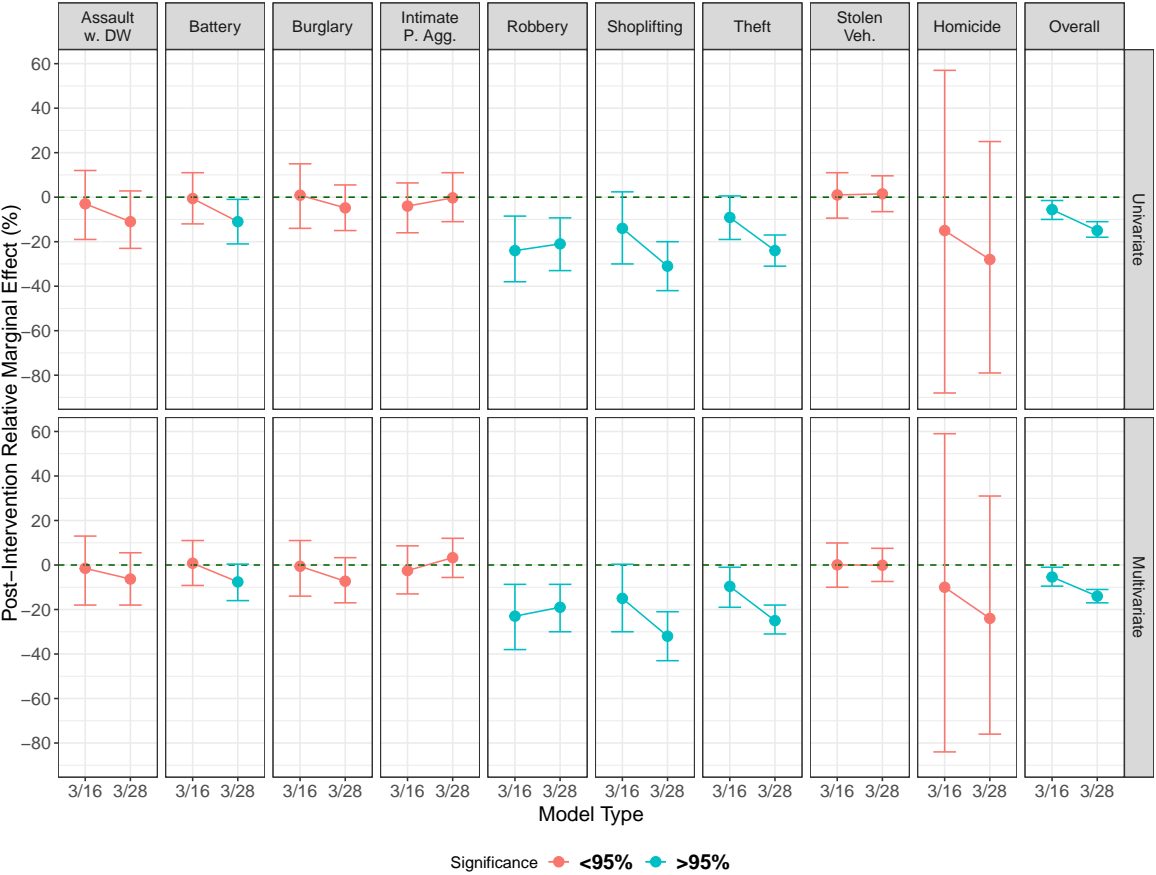}
    \caption{Graphical Summary of Model Results (with 95\% Confidence Intervals)}
    \label{fig:summaryline}
\end{figure}

Finally, we considered all reported crimes in the two selected temporal windows under analysis; this aggregated variable also accounts for all those offenses that go beyond the categories previously considered. Figure \ref{fig:summaryline} displays the evolution of the post intervention relative marginal effect in the two considered time windows, for all crimes and model types. In particular, the graph represents the percent changes in crime between our synthetic control, which we set at zero, and the actual registered crimes. Our empirical results indicate a significant decrease in the overall crime occurrence in the first weeks after the introduction of social distancing measures compared to the virtual counterfactual scenario with no intervention. In particular, the models focusing on the mild policies period show a reduction of 5.6\% and 5.4\% of overall crimes in the univariate and multivariate cases, respectively. In the models considering the longer time-window when strict policies were also considered, instead, the decrease reaches -15\% and -14\% in the two cases. The posterior probability of a causal effect is equal to 99.4\% and 98.5\% for the first temporal window and 99.7\% and 99.8\% for the second temporal window, showing a high level of significance.

When focusing on all reported crimes, the results are in line with those seen for robbery, shoplifting, theft, and battery. Nonetheless, the cumulated count of these crimes cannot explain on its own the overall decrease given the numerosity of other crimes. As such, these findings suggest that, potentially, most crimes have diminished in the considered short-term period leading to a general reduction of crime in the entire city of Los Angeles. In turn, this result indicates that, besides the categories chosen for this study, further work is needed to disentangle mechanisms related to single offenses or crime categories.

\section{Discussion \& Conclusion} \label{dicussion}
In line with our hypotheses, the statistical results show that robberies, thefts, and shoplifting had a statistically significant reduction already in the post-intervention period up to March 16\textsuperscript{th}, when only mild policies were applied. Robbery recorded the single largest decrease (-23\%/-24\% depending on the statistical specifications). \cite{MohlerImpactsocialdistancing2020}, applying a different statistical strategy on different data and concentrating on a partially different time frame, also observed a significant reduction in robberies in Los Angeles in the immediate aftermath of the introduction of social-distancing measures. After robberies, shoplifting (-14\%/-15\%) and thefts (-9.1\%/-9.6\%) are the crime categories for which the differences between the observed data and the virtual scenario are the strongest. Overall crimes also significantly declined (-5.4\%/-5.6\%). Theft, robbery and shoplifting are offenses involving a direct act targeting a property, thus making it likely that common underlying principles govern the similar decreasing trends of these crimes. These reductions can be explained in light of the reduction of social interactions as people avoid public spaces and spend more time at home as well as by closure of public places and the presence of quotas on the number of entries into shops and malls. In turn, this leads to a reduction in criminal opportunities and to an increase in guardianship.
Coherently with the ideas emerging from routine activity theory \citep{CohenSocialChangeCrime1979} and crime pattern theory \citep{BrantinghamPatternscrime1984}, most crime reductions are larger after the introduction of stricter containment policies, which further reduce criminal opportunities while increasing informal guardianship. The models considering the entire period up to March 28\textsuperscript{th} show a significant reduction for shoplifting (-31\%/-32\%), thefts (-24\%/-25\%), robberies (-21\%/-19\%), and for overall crimes (-15\%/-14\%). In addition, battery (-11\%/7.6\%) also started to show a significant decrease after the adoption of strict distancing policies.

Contrarily, the models did not detect any significant change in the trends of homicides, assaults with a deadly weapon, intimate partner assaults, but also of burglaries and stolen vehicles. Previous studies have listed several stressors specifically emerging during quarantines comprising the duration of the quarantine itself, the fear of infection, frustration and boredom, and inadequate supply of basic commodities, services, and information \citep{Brookspsychologicalimpactquarantine2020}. In addition, families under financial and psychological stress as a result of the pandemic, increased their alcohol use at home \citep{ColbertCOVID19alcohol2020}. This might also increase the occurrence of violent and expressive crimes since there is a strong relationship between the use and abuse of alcohol and violence \citep{ParkerAlcoholDrugsViolence1998}. In line with these evidence on the negative psychological outcomes of quarantine, our statistical results suggest that besides reduction of opportunities and change in social life, increased strain is likely to have a leading role in explaining crime dynamics immediately after the introduction of social distancing measures. In this perspective, the lack of significant variations in the trends of the aforementioned crimes might be interpreted as the result of a combined effect of a partial reduction in situational and opportunistic triggers of crime, on the one hand, and worsening of the balance between positive and negative psychological stimuli on the other.

In light of this, intimate partner assault warrants some further considerations. Firstly, intimate partner assault is a crime which is strongly related to strain dynamics \citep{Erikssongeneralstraintheory2013, PiqueroStayingHomeStaying2020}. Secondly, crime factors pertaining to routine activity theory and crime pattern theory and factors pertaining to general strain theory all suggest an increase in intimate partner assaults. It is not so for other considered crimes, with the partial exception of vehicle theft. Nonetheless, we do not observe any statistically significant increase in the count of intimate partner assaults. By contrast, \cite{MohlerImpactsocialdistancing2020}, in their study, saw significant increases in domestic violence calls for service in Los Angeles. In this regard, to be noted is that many of such calls concern domestic disturbances without a physical assault \citep{MacDonaldPoliceuseforce2003, MohlerImpactsocialdistancing2020}.

More in general, our analysis suggests that containment policies in the immediate aftermath have had a stronger impact on more instrumental crimes compared to crimes for which expressive motivations are more relevant. Similarly, more serious offenses appear to be less influenced by social-distancing measures. The combination of motivation and seriousness identifies two macro classes of crimes, which may help the interpretation of the effects of containment policies. More instrumental/less serious crimes are strongly influenced by the reduction of social interactions and the sudden modification of everyday habits. While routine activity theory and crime pattern theory seem to be a powerful tool to interpret changes in this first class of criminal behaviors, they are less effective in capturing the dynamics of more expressive/more serious crimes whose interpretation needs to be integrated with a broader set of factors \citep{HaywardSituationalCrimePrevention2007}. Strong motivation, low self-control and criminal opportunities are concurrent causes in the explanation of crime commitment in the aftermath of the policy introduction, but the combination of these causes is not equal for all crime types \citep{LongshoreSelfControlCriminalOpportunity1998}. It appears that strong motivation and low self-control are more likely to play a central role in more serious crimes, whereas criminal opportunities do so in triggering less serious offenses. In this regard, the strain theory of crime \citep{AgnewFoundationGeneralStrain1992} may act as complementary interpretative framework to the rational choice approach, explaining the emerging trends also in light of the sudden amplification of stressors resulting from COVID-19 containment measures. Future studies may investigate these dynamic more in depth by exploiting a panel approach and by identifying variables capable of representing the different theories we exploited in our interpretative framework.

Finally, burglaries and vehicle thefts appear to fall outside this reasoning and possibly nullify it. Yet, as previously said, these are crime categories which unify different actual criminal dynamics. Burglaries can be residential or non-residential; separating the two may shed light on distinct opposite processes, as residential burglaries are expected to decrease, while non-residential ones to increase as a byproduct of different dynamics in guardianship. In support of this, \cite{AshbyInitialevidencerelationship2020}, who could separate residential and non-residential burglaries in Los Angeles, observes–at least during some weeks–significant changes with respect to both residential and non-residential burglaries. 
It is thus important to mention again that our results capture short-term dynamics within an open time-frame that will likely extend over several weeks ahead, thus potentially leading to new patterns. Furthermore, we may have witnessed, especially in relation to the period up to March 16\textsuperscript{th}, a mixed dynamic combining both the effect of the mild policies and the fear of contagion. Fear of contagion has probably played a role in reshaping people’s individual and collective behavior \citep{Brookspsychologicalimpactquarantine2020}, even before the second half of March when stronger measures were adopted. This process, directly impacting on the density and frequency of social contacts, has potentially fostered the effect of the policy recommendations on crime.

In addition, social distance restrictions may have also influenced the reporting rate of criminal offenses. People avoid spending time outside their homes, and this may reduce their willingness or ability to go to the police to report a crime. In the case of Los Angeles, problems related to a reduction of crime reporting are mitigated by the possibility to report a crime using an online form for specific crimes (e.g., theft or theft from vehicles) or calling a dedicated number. Moreover, while underreporting issues may affect the count of all the crimes considered, the crimes for which we do not observe any significant change–with the important exception of intimate partner violence–are the ones that are more likely to be reported–i.e., homicide, assaults with deadly weapons, stolen vehicle, and burglary. These factors help to reduce possible biases due to increases in underreporting.

By contrast, the dynamics related to the lack of–or impossibility to–reporting may especially influence the count of intimate partner assaults. Indeed, the cohabitation of victim and offender may make it difficult for the victim to report the offense to the police. Nevertheless, since neighbors are likewise at home, they are also better able to exert partial guardianship on the episodes of intimate partner violence that are happening in their surroundings and call the police. In this regard, studies have demonstrated that neighbors can play a role in reporting abuses \citep{PaquinStatewideSurveyReactions1994}. In addition, scholars have shown that informal actors (e.g., family, friends, and neighbors) are often approached by the victims of intimate partner violence because they are proximal and may be able to intervene before, during, and after the violent event \citep{McCartHelpseekingvictims2010, WeeModifiersNeighborsBystander2016}. Related to this, the police attention to urban crimes may have diminished due to the need to enforce social distancing measures, thus partially countering the crime mitigating effect driven by looser social interactions. However, the reduction of police capabilities is likely to have only a only marginal effect on intimate partner assaults, which are serious crimes that often takes place in a private environment.

Additional caution should be adopted in interpreting the results obtained. As the extant criminological literature indicates, crime clusters in time and space, meaning that its  spatio-temporal distribution is not random \citep{DoranInvestigatingSpatiotemporalLinks2005, GrubesicSpatioTemporalInteractionUrban2008, MohlerSelfExcitingPointProcess2011, WeisburdLawCrimeConcentration2015}. Therefore, there are very few reasons to think that this patterned nature will entirely change because of containment measures. Future work will hence need to investigate the potential heterogeneity of policy effects across different areas of the same city, also analyzing the potential correlates of diverging trends in terms environmental and socio-economic factors. Furthermore, according to anecdotal evidence, police were asked to limit or stop making low-level arrests to manage jail crowding and narrow down the spread of the virus in prisons. These changes in police practice could have influenced crime trends and we could expect a downward bias for the estimates of less serious crimes (e.g., shoplifting). The same downward bias cannot be foreseen for homicide or intimate partner assaults that are not low-level crimes, but serious offences instead. An opposite trend that could have affected crime rates in the opposite direction in the period under analysis is the early release of prisoners in order to contrast the spread of coronavirus within prisons. The early release of prisoners could potentially lead to an increase in crime rates due to recidivist behaviors of former inmates. The two mechanisms described here go in opposite directions (decrease vs increase in crime trends), therefore they potentially tend to overall balance themselves. 

\section{Policy Propositions}

This work has outlined how COVID-19 containment policies have influenced criminal trends in the city of Los Angeles already in the first weeks after their introduction. Policy implications emerge from these analyses. First, studying criminal dynamics in this anomalous time should help in creating protective measures for the most vulnerable subjects influenced by these changes, including homeless people, women and children. Homeless people, for instance, could experience an increase in the likelihood of becoming victims of crimes, as a consequence of lower guardianship and social control. Furthermore, as people are likely to suffer more stress in this period thus being more prone to commit certain crimes, including violent ones, it might be worth empowering remote contact points, facilitating distance reporting to make it easier for victims, who cannot freely move, to connect with the police. Distance reporting, for instance, could include text messaging services to reduce the risk of being stopped or heard by the abusive member of a household.

Finally, as crime takes new forms and dynamics, law enforcement agencies will be required to modify or re-define resource allocation for the new priorities. The task, especially in highly populated and heterogeneous cities like Los Angeles, can be herculean. In fact, while many police departments and other institutions in the United States have tailored their actions based on predictive policing software, these models can suddenly become of little help. Depending on the extent to which these policies will force crime to change–beyond mere temporal trends–predictive models built on millions of past observations may no longer be informative. This situation thus urges alternative predictive tools that can take into account disruptions of social life as the triggers of new criminal risks, prompting data-driven strategies to re-assess criminal patterns and countering strategies.

\section*{Conflict of Interest}
The authors declare no conflict of interest.
\section*{Funding}
The authors did not receive any funding for the present work.

\section*{Acknowledgments}
The authors wish to thank Jay Aronson, Laura Dugan, Marco Dugato, Gary LaFree, Clarissa Manning, Cecilia Meneghini, Riccardo Milani, James Prieger and two anonymous referees for their precious feedback and comments on previous versions of this manuscript. 

\newpage
\bibliographystyle{elsarticle-num-names}
\bibliography{covid_epidemics_lit.bib}

\newpage
\section*{Supplementary Materials}
\setcounter{table}{0}
\renewcommand{\thetable}{A\arabic{table}}

\begin{table}[!hbt]
\centering
\footnotesize
\begin{tabular}{lcccc}
\hline
 & \multicolumn{4}{c}{\textbf{Daily crime counts up to March 16\textsuperscript{th}}} \\ 
\textbf{} & \multicolumn{2}{c}{\textbf{Univariate}} & \multicolumn{2}{c}{\textbf{With Cov.}} \\ 
 & \textbf{Avg.} & \textbf{Cum.} & \textbf{Avg.} & \textbf{Cum.} \\ \hline
Actual & 26 & 317 & 26 & 317 \\
Prediction (S.D.) & 27 (2.2) & 327 (26.9) & 27 (2.2) & 327 (27.1) \\
95\% C.I. & {[}23,32{]} & {[}275,379{]} & {[}23,31{]} & {[}274, 374{]} \\ \hline
Absolute Effect (S.D.) & -0.79 (2.2) & -9.50 (26.9) & -0.41 (2.3) & -4.89 (27.1) \\
95\% C.I. & {[}-5.1, 3.5{]} & {[}-61.6, 42.2{]} & {[}-4.7, 3.6{]} & {[}-56.9, 43.1{]} \\ \hline
Relative Effect (S.D.) & -2.98\% (8.2\%) & -2.98\% (8.2\%) & -1.5\% (8.4\%) & -1.5\% (8.4\%) \\
95\% C.I. & {[}-19\%, 13\%{]} & {[}-19\%, 13\%{]} & {[}-18\%, 13\%{]} & {[}-18\%, 13\%{]} \\ \hline
Post. tail-area prob. p: & \multicolumn{2}{c}{0.36409} & \multicolumn{2}{c}{0.48134} \\
Post. prob. causal effect: & \multicolumn{2}{c}{64\%} & \multicolumn{2}{c}{52\%} \\ \hline\hline
 & \multicolumn{4}{c}{\textbf{Daily crime counts up to March 28\textsuperscript{th}}} \\ 
 & \multicolumn{2}{c}{\textbf{Univariate}} & \multicolumn{2}{c}{\textbf{With Cov.}} \\ 
 & \textbf{Avg.} & \textbf{Cum.} & \textbf{Avg.} & \textbf{Cum.} \\ \hline
Actual & 24 & 604 & 24 & 604 \\
Prediction (S.D.) & 27 (1.8) & 675 (45.7) & 26 (1.5) & 644 (38.4) \\
95\% C.I. & {[}23, 30{]} & {[}585, 760{]} & {[}23, 29{]} & {[}568, 718{]} \\ \hline
Absolute Effect (S.D.) & -2.9 (1.8) & -71.3 (45.7) & -1.6 (1.5) & -40.4 (38.4) \\
95\% C.I. & {[}-6.2, 0.76{]} & {[}-155.6, 19.00{]} & {[}-4.6, 1.4{]} & {[}-114.2, 35.6{]} \\ \hline
Relative Effect (S.D.) & -11\% (6.8\%) & -11\% (6.8\%) & -6.3\% (6\%) & -6.3\% (6\%) \\
95\% C.I. & {[}-23\%, 2.8\%{]} & {[}-23\%, 2.8\%{]} & {[}-18\%, 5.5\%{]} & {[}-18\%, 5.5\%{]} \\ \hline
Post. tail-area prob. p: & \multicolumn{2}{c}{0.0625} & \multicolumn{2}{c}{0.14801} \\
Post. prob. causal effect: & \multicolumn{2}{c}{94\%} & \multicolumn{2}{c}{85\%} \\ \hline
\end{tabular}
\caption{Causal Impact Analysis - Assaults with Deadly Weapons}
\label{assaults}
\end{table}

\begin{table}[!hbt]
\centering
\footnotesize
\begin{tabular}{lcccc}
\hline
 & \multicolumn{4}{c}{\textbf{Daily crime counts up to March 16\textsuperscript{th}}} \\
 & \multicolumn{2}{c}{\textbf{Univariate}} & \multicolumn{2}{c}{\textbf{With Cov.}} \\
 & \textbf{Average} & \textbf{Cumulative} & \textbf{Average} & \textbf{Cumulative} \\ \hline
Actual & 50 & 597 & 50 & 597 \\
Prediction (S.D.) & 50 (2.9) & 601 (34.4) & 49 (2.7) & 592 (32.5) \\
95\% C.I. & {[}44, 56{]} & {[}533, 668{]} & {[}44, 54{]} & {[}530, 652{]} \\ \hline
Absolute Effect (S.D.) & -0.3 (2.9) & -3.6 (34.4) & 0.39 (2.7) & 4.62 (32.5) \\
95\% C.I. & {[}-5.9, 5.3{]} & {[}-71.2, 64.1{]} & {[}-4.5, 5.6{]} & {[}-54.5, 66.7{]} \\ \hline
Relative Effect (S.D.) & -0.6\% (5.7\%) & -0.6\% (5.7\%) & 0.78\% (5.5\%) & 0.78\% (5.5\%) \\
95\% C.I. & {[}-12\%, 11\%{]} & {[}-12\%, 11\%{]} & {[}-9.2\%, 11\%{]} & {[}-9.2\%, 11\%{]} \\ \hline
Post. tail-area prob. p: & \multicolumn{2}{c}{0.41895} & \multicolumn{2}{c}{0.40672} \\
Post. prob. causal effect: & \multicolumn{2}{c}{58\%} & \multicolumn{2}{c}{59\%} \\ \hline\hline
 & \multicolumn{4}{c}{\textbf{Daily crime counts up to March 28\textsuperscript{th}}} \\
 & \multicolumn{2}{c}{\textbf{Univariate}} & \multicolumn{2}{c}{\textbf{With Cov.}} \\
 & \textbf{Avg.} & \textbf{Cum.} & \textbf{Avg.} & \textbf{Cum.} \\ \hline
Actual & 45 & 1116 & 45 & 1116 \\
Prediction (S.D.) & 50 (2.5) & 1256 (63.1) & 48 (2) & 1207 (50) \\
95\% C.I. & {[}45, 55{]} & {[}1128, 1380{]} & {[}44, 52{]} & {[}1111, 1307{]} \\ \hline
Absolute Effect (S.D.) & -5.6 (2.5) & -139.7 (63.1) & -3.7 (2) & -91.3 (50) \\
95\% C.I. & {[}-11, -0.5{]} & {[}-264, -12.4{]} & {[}-7.6, 0.19{]} & {[}-191.1, 4.74{]} \\ \hline
Relative Effect (S.D.) & -11\% (5\%) & -11\% (5\%) & -7.6\% (4.1\%) & -7.6\% (4.1\%) \\
95\% C.I. & {[}-21\%, -0.99\%{]} & {[}-21\%, -0.99\%{]} & {[}-16\%, 0.39\%{]} & {[}-16\%, 0.39\%{]} \\ \hline
Post. tail-area prob. p: & \multicolumn{2}{c}{0.02486} & \multicolumn{2}{c}{0.03534} \\
Post. prob. causal effect: & \multicolumn{2}{c}{98\%} & \multicolumn{2}{c}{96\%} \\ \hline
\end{tabular}
\caption{Causal Impact Analysis - Battery and Simple Assault}
\label{battery}
\end{table}

\begin{table}[!hbt]
\centering
\footnotesize
\begin{tabular}{lcccc}
\hline
 & \multicolumn{4}{c}{\textbf{Daily crime counts up to March 16\textsuperscript{th}}} \\
 & \multicolumn{2}{c}{\textbf{Univariate}} & \multicolumn{2}{c}{\textbf{With Cov.}} \\
 & \textbf{Average} & \textbf{Cumulative} & \textbf{Average} & \textbf{Cumulative} \\ \hline\hline
Actual & 34 & 406 & 34 & 406 \\
Prediction (S.D.) & 34 (2.4) & 402 (29.2) & 34 (2.5) & 408 (29.6) \\
95\% C.I. & {[}29, 38{]} & {[}345, 461{]} & {[}30, 39{]} & {[}359, 465{]} \\ \hline
Absolute Effect (S.D.) & 0.3 (2.4) & 3.6 (29.2) & -0.2 (2.5) & -2.4 (29.6) \\
95\% C.I. & {[}-4.6, 5.1{]} & {[}-54.9, 60.8{]} & {[}-4.9, 3.9{]} & {[}-59.2, 46.6{]} \\ \hline
Relative Effect (S.D.) & 0.89\% (7.2\%) & 0.89\% (7.2\%) & -0.58\% (7.2\%) & -0.58\% (7.2\%) \\
95\% C.I. & {[}-14\%, 15\%{]} & {[}-14\%, 15\%{]} & {[}-14\%, 11\%{]} & {[}-14\%, 11\%{]} \\ \hline
Post. tail-area prob. p: & \multicolumn{2}{c}{0.44173} & \multicolumn{2}{c}{0.49627} \\
Post. prob. causal effect: & \multicolumn{2}{c}{56\%} & \multicolumn{2}{c}{50\%} \\ \hline\hline
& \multicolumn{4}{c}{\textbf{Daily crime counts up to March 28\textsuperscript{th}}} \\
& \multicolumn{2}{c}{\textbf{Univariate}} & \multicolumn{2}{c}{\textbf{With Cov.}} \\
& \textbf{Avg.} & \textbf{Cum.} & \textbf{Avg.} & \textbf{Cum.} \\ \hline
Actual & 33 & 816 & 33 & 816 \\
Prediction (S.D.) & 34 (1.9) & 858 (46.3) & 35 (1.8) & 881 (46.1) \\
95\% C.I. & {[}31, 38{]} & {[}769, 948{]} & {[}31, 39{]} & {[}787, 966{]} \\ \hline
Absolute Effect (S.D.) & -1.7 (1.9) & -41.6 (46.3) & -2.6 (1.8) & -64.5 (46.1) \\
95\% C.I. & {[}-5.3, 1.9{]} & {[}-131.7, 47.1{]} & {[}-6, 1.2{]} & {[}-150, 29.3{]} \\ \hline
Relative Effect (S.D.) & -4.8\% (5.4\%) & -4.8\% (5.4\%) & -7.3\% (5.2\%) & -7.3\% (5.2\%) \\
95\% C.I. & {[}-15\%, 5.5\%{]} & {[}-15\%, 5.5\%{]} & {[}-17\%, 3.3\%{]} & {[}-17\%, 3.3\%{]} \\ \hline
Post. tail-area prob. p: & \multicolumn{2}{c}{0.19832} & \multicolumn{2}{c}{0.09328} \\
Post. prob. causal effect: & \multicolumn{2}{c}{80\%} & \multicolumn{2}{c}{91\%} \\ \hline
\end{tabular}
\caption{Causal Impact Analysis - Burglary}
\label{burglary}
\end{table}

\begin{table}[!hbt]
\centering
\footnotesize
\begin{tabular}{lcccc}
\hline
 & \multicolumn{4}{c}{\textbf{Daily crime counts up to March 16\textsuperscript{th}}} \\
 & \multicolumn{2}{c}{\textbf{Univariate}} & \multicolumn{2}{c}{\textbf{With Cov.}} \\ 
 & \textbf{Average} & \textbf{Cumulative} & \textbf{Average} & \textbf{Cumulative} \\ \hline\hline
Actual & 38 & 454 & 38 & 454 \\
Prediction (S.D.) & 39 (2.3) & 473 (27.5) & 39 (2.1) & 466 (25.3) \\
95\% C.I. & {[}35, 44{]} & {[}424, 529{]} & {[}34, 43{]} & {[}414, 512{]} \\ \hline
Absolute Effect (S.D.) & -1.6 (2.3) & -18.8 (27.5) & -0.96 (2.1) & -11.52 (25.3) \\
95\% C.I. & {[}-6.3, 2.5{]} & {[}-75.4, 30.3{]} & {[}-4.9, 3.4{]} & {[}-58.4, 40.2{]} \\ \hline
Relative Effect (S.D.) & -4\% (5.8\%) & -4\% (5.8\%) & -2.5\% (5.4\%) & -2.5\% (5.4\%) \\
95\% C.I. & {[}-16\%, 6.4\%{]} & {[}-16\%, 6.4\%{]} & {[}-13\%, 8.6\%{]} & {[}-13\%, 8.6\%{]} \\ \hline
Post. tail-area prob. p: & \multicolumn{2}{c}{0.24535} & \multicolumn{2}{c}{0.37313} \\
Post. prob. causal effect: & \multicolumn{2}{c}{75\%} & \multicolumn{2}{c}{63\%} \\ \hline\hline
 & \multicolumn{4}{c}{\textbf{Daily crime counts up to March 28\textsuperscript{th}}} \\
 & \multicolumn{2}{c}{\textbf{Univariate}} & \multicolumn{2}{c}{\textbf{With Cov.}} \\
 & \textbf{Avg.} & \textbf{Cum.} & \textbf{Avg.} & \textbf{Cum.} \\ \hline
Actual & 39 & 969 & 39 & 969 \\
Prediction (S.D.) & 39 (2.2) & 972 (56.0) & 38 (1.7) & 938 (41.7) \\
95\% C.I. & {[}34, 43{]} & {[}860, 1071{]} & {[}34, 41{]} & {[}855, 1021{]} \\ \hline
Absolute Effect (S.D.) & -0.11 (2.2) & -2.68 (56.0) & 1.2 (1.7) & 30.9 (41.7) \\
95\% C.I. & {[}-4.1, 4.4{]} & {[}-102.3, 109.3{]} & {[}-2.1, 4.5{]} & {[}-52.1, 113.7{]} \\ \hline
Relative Effect (S.D.) & -0.28\% (5.8\%) & -0.28\% (5.8\%) & 3.3\% (4.4\%) & 3.3\% (4.4\%) \\
95\% C.I. & {[}-11\%, 11\%{]} & {[}-11\%, 11\%{]} & {[}-5.6\%, 12\%{]} & {[}-5.6\%, 12\%{]} \\ \hline
Post. tail-area prob. p: & \multicolumn{2}{c}{0.48324} & \multicolumn{2}{c}{0.22015} \\
Post. prob. causal effect: & \multicolumn{2}{c}{52\%} & \multicolumn{2}{c}{78\%} \\ \hline
\end{tabular}
\caption{Causal Impact Analysis - Intimate Partner Assault}
\label{intimate}
\end{table}

\begin{table}[!hbt]
\centering
\footnotesize
\begin{tabular}{lcccc}
\hline
 & \multicolumn{4}{c}{\textbf{Daily crime counts up to March 16\textsuperscript{th}}} \\
 & \multicolumn{2}{c}{\textbf{Univariate}} & \multicolumn{2}{c}{\textbf{With Cov.}} \\ \hline
 & \textbf{Avg.} & \textbf{Cum.} & \textbf{Avg.} & \textbf{Cum.} \\ \hline
Actual & 17 & 202 & 17 & 202 \\
Prediction (S.D.) & 22 (1.7) & 266 (20.5) & 22 (1.6) & 262 (19.5) \\
95\% C.I. & {[}19, 25{]} & {[}225, 304{]} & {[}19, 25{]} & {[}225, 302{]} \\ \hline
Absolute Effect (S.D.) & -5.3 (1.7) & -63.6 (20.5) & -5 (1.6) & -60 (19.5) \\
95\% C.I. & {[}-8.5, -1.9{]} & {[}-101.8, -22.6{]} & {[}-8.3, -1.9{]} & {[}-100.1, -22.8{]} \\ \hline
Relative Effect (S.D.) & -24\% (7.7\%) & -24\% (7.7\%) & -23\% (7.5\%) & -23\% (7.5\%) \\
95\% C.I. & {[}-38\%, -8.5\%{]} & {[}-38\%, -8.5\%{]} & {[}-38\%, -8.7\%{]} & {[}-38\%, -8.7\%{]} \\ \hline
Post. tail-area prob. p: & \multicolumn{2}{c}{0.00333} & \multicolumn{2}{c}{0.00208} \\
Post. prob. causal effect: & \multicolumn{2}{c}{99.67\%} & \multicolumn{2}{c}{99.79\%} \\ \hline\hline
 & \multicolumn{4}{c}{\textbf{Daily crime counts up to March 28\textsuperscript{th}}} \\
 & \multicolumn{2}{c}{\textbf{Univariate}} & \multicolumn{2}{c}{\textbf{With Cov.}} \\
 & \textbf{Avg.} & \textbf{Cum.} & \textbf{Avg.} & \textbf{Cum.} \\ \hline
Actual & 18 & 439 & 18 & 439 \\
Prediction (S.D.) & 22 (1.3) & 553 (32.8) & 22 (1.2) & 545 (30.4) \\
95\% C.I. & {[}20, 25{]} & {[}490, 619{]} & {[}19, 24{]} & {[}486, 604{]} \\ \hline
Absolute Effect (S.D.) & -4.6 (1.3) & -114.2 (32.8) & -4.3 (1.2) & -106.3 (30.4) \\
95\% C.I. & {[}-7.2, -2.1{]} & {[}-180.5, -51.3{]} & {[}-6.6, -1.9{]} & {[}-165.4, -47.4{]} \\ \hline
Relative Effect (S.D.) & -21\% (5.9\%) & -21\% (5.9\%) & -19\% (5.6\%) & -19\% (5.6\%) \\
95\% C.I. & {[}-33\%, -9.3\%{]} & {[}-33\%, -9.3\%{]} & {[}-30\%, -8.7\%{]} & {[}-30\%, -8.7\%{]} \\ \hline
Post. tail-area prob. p: & \multicolumn{2}{c}{0.00298} & \multicolumn{2}{c}{0.00218} \\
Post. prob. causal effect: & \multicolumn{2}{c}{99.70\%} & \multicolumn{2}{c}{99.78\%} \\ \hline
\end{tabular}
\caption{Causal Impact Analysis - Robbery}
\label{robbery}
\end{table}

\begin{table}[!hbt]
\centering
\footnotesize
\begin{tabular}{lcccc}
\hline
 & \multicolumn{4}{c}{\textbf{Daily crime counts up to March 16\textsuperscript{th}}} \\
 & \multicolumn{2}{c}{\textbf{Univariate}} & \multicolumn{2}{c}{\textbf{With Cov.}} \\ \hline
 & \textbf{Avg.} & \textbf{Cum.} & \textbf{Avg.} & \textbf{Cum.} \\ \hline
Actual & 16 & 189 & 16 & 189 \\
Prediction (S.D.) & 18 (1.5) & 220 (18.2) & 19 (1.5) & 223 (17.7) \\
95\% C.I. & {[}15, 21{]} & {[}184, 255{]} & {[}16, 21{]} & {[}188, 256{]} \\ \hline
Absolute Effect (S.D.) & -2.6 (1.5) & -31.1 (18.2) & -2.8 (1.5) & -34.1 (17.7) \\
95\% C.I. & {[}-5.5, 0.44{]} & {[}-66.3, 5.30{]} & {[}-5.6, 0.064{]} & {[}-67.4, 0.766{]} \\ \hline
Relative Effect (S.D.) & -14\% (8.3\%) & -14\% (8.3\%) & -15\% (7.9\%) & -15\% (7.9\%) \\
95\% C.I. & {[}-30\%, 2.4\%{]} & {[}-30\%, 2.4\%{]} & {[}-30\%, 0.34\%{]} & {[}-30\%, 0.34\%{]} \\ \hline
Post. tail-area prob. p: & \multicolumn{2}{c}{0.04353} & \multicolumn{2}{c}{0.02808} \\
Post. prob. causal effect: & \multicolumn{2}{c}{95.64\%} & \multicolumn{2}{c}{97.19\%} \\\hline\hline
 & \multicolumn{4}{c}{\textbf{Daily crime counts up to March 28\textsuperscript{th}}} \\
 & \multicolumn{2}{c}{\textbf{Univariate}} & \multicolumn{2}{c}{\textbf{With Cov.}} \\
 & \textbf{Avg.} & \textbf{Cum.} & \textbf{Avg.} & \textbf{Cum.} \\ \hline
Actual & 13 & 320 & 13 & 320 \\
Prediction (S.D.) & 18 (1) & 462 (26) & 19 (1) & 471 (26) \\
95\% C.I. & {[}16, 21{]} & {[}412, 513{]} & {[}17, 21{]} & {[}421, 521{]} \\ \hline
Absolute Effect (S.D.) & -5.7 (1) & -142.0 (26) & -6.1 (1) & -151.5 (26) \\
95\% C.I. & {[}-7.7, -3.7{]} & {[}-193.4, -92.0{]} & {[}-8, -4.1{]} & {[}-201, -101.3{]} \\ \hline
Relative Effect (S.D.) & -31\% (5.6\%) & -31\% (5.6\%) & -32\% (5.5\%) & -32\% (5.5\%) \\
95\% C.I. & {[}-42\%, -20\%{]} & {[}-42\%, -20\%{]} & {[}-43\%, -21\%{]} & {[}-43\%, -21\%{]} \\ \hline
Post. tail-area prob. p: & \multicolumn{2}{c}{0.00348} & \multicolumn{2}{c}{0.001} \\
Post. prob. causal effect: & \multicolumn{2}{c}{99.65\%} & \multicolumn{2}{c}{99.90\%} \\
\hline\hline

\end{tabular}
\caption{Causal Impact Analysis - Shoplifting}
\label{shoplifting}
\end{table}

\begin{table}[!hbt]
\centering
\footnotesize
\begin{tabular}{lcccc}
\hline
 & \multicolumn{4}{c}{\textbf{Daily crime counts up to March 16\textsuperscript{th}}} \\
 & \multicolumn{2}{c}{\textbf{Univariate}} & \multicolumn{2}{c}{\textbf{With Cov.}} \\
 & \textbf{Avg.} & \textbf{Cum.} & \textbf{Avg.} & \textbf{Cum.} \\ \hline
Actual & 55 & 662 & 55 & 662 \\
Prediction (S.D.) & 61 (2.9) & 728 (35.3) & 61 (2.8) & 732 (33.2) \\
95\% C.I. & {[}55, 66{]} & {[}658, 797{]} & {[}56, 67{]} & {[}670, 801{]} \\ \hline
Absolute Effect (S.D.) & -5.5 (2.9) & -66.1 (35.3) & -5.8 (2.8) & -70.0 (33.2) \\
95\% C.I. & {[}-11, 0.34{]} & {[}-135, 4.13{]} & {[}-12, -0.64{]} & {[}-139, -7.67{]} \\ \hline
Relative Effect (S.D.) & -9.1\% (4.8\%) & -9.1\% (4.8\%) & -9.6\% (4.5\%) & -9.6\% (4.5\%) \\
95\% C.I. & {[}-19\%, 0.57\%{]} & {[}-19\%, 0.57\%{]} & {[}-19\%, -1\%{]} & {[}-19\%, -1\%{]}\\\hline
Posterior tail-area probability p: & \multicolumn{2}{c}{0.0333} & \multicolumn{2}{c}{0.01663} \\
Posterior prob. of a causal effect: & \multicolumn{2}{c}{96.67\%} & \multicolumn{2}{c}{98.33\%} \\\hline\hline
 & \multicolumn{4}{c}{\textbf{Daily crime counts up to March 28\textsuperscript{th}}} \\
 & \multicolumn{2}{c}{\textbf{Univariate}} & \multicolumn{2}{c}{\textbf{With Cov.}} \\
 & \textbf{Avg.} & \textbf{Cum.} & \textbf{Avg.} & \textbf{Cum.} \\ \hline
Actual & 47 & 1175 & 47 & 1175 \\
Prediction (S.D.) & 62 (2.2) & 1548 (56.2) & 62 (2.1) & 1557 (53.4) \\
95\% C.I. & {[}58, 66{]} & {[}1444, 1651{]} & {[}58, 66{]} & {[}1453, 1658{]} \\ \hline
Absolute Effect (S.D.) & -15 (2.2) & -373 (56.2) & -15 (2.1) & -382 (53.4) \\
95\% C.I. & {[}-19, -11{]} & {[}-476, -269{]} & {[}-19, -11{]} & {[}-483, -278{]} \\ \hline
Relative Effect (S.D.) & -24\% (3.6\%) & -24\% (3.6\%) & -25\% (3.4\%) & -25\% (3.4\%) \\
95\% C.I. & {[}-31\%, -17\%{]} & {[}-31\%, -17\%{]} & {[}-31\%, -18\%{]} & {[}-31\%, -18\%{]} \\ \hline
Post. tail-area prob. p: & \multicolumn{2}{c}{0.0035} & \multicolumn{2}{c}{0.00109} \\
Post. prob. causal effect: & \multicolumn{2}{c}{99.65\%} & \multicolumn{2}{c}{99.89\%} \\ \hline
\end{tabular}
\caption{Causal Impact Analysis - Thefts}
\label{thefts}
\end{table}

\begin{table}[!hbt]
\centering
\footnotesize
\begin{tabular}{lcccc}
\hline
 & \multicolumn{4}{c}{\textbf{Daily crime counts up to March 16\textsuperscript{th}}} \\
 & \multicolumn{2}{c}{\textbf{Univariate}} & \multicolumn{2}{c}{\textbf{With Cov.}} \\ \hline
 & \textbf{Avg.} & \textbf{Cum.} & \textbf{Avg.} & \textbf{Cum.} \\ \hline
Actual & 45 & 536 & 45 & 536 \\
Prediction (S.D.) & 44 (2.4) & 530 (28.6) & 45 (2.4) & 536 (28.8) \\
95\% C.I. & {[}40, 49{]} & {[}477, 586{]} & {[}40, 49{]} & {[}483, 591{]} \\ \hline
Absolute Effect (S.D.) & 0.46 (2.4) & 5.55 (28.6) & 0.029 (2.4) & 0.343 (28.8) \\
95\% C.I. & {[}-4.2, 4.9{]} & {[}-49.8, 58.8{]} & {[}-4.6, 4.4{]} & {[}-55.1, 52.8{]} \\ \hline
Relative Effect (S.D.) & 1\% (5.4\%) & 1\% (5.4\%) & 0.064\% (5.4\%) & 0.064\% (5.4\%) \\
95\% C.I. & {[}-9.4\%, 11\%{]} & {[}-9.4\%, 11\%{]} & {[}-10\%, 9.9\%{]} & {[}-10\%, 9.9\%{]} \\ \hline
Post. tail-area prob. p: & \multicolumn{2}{c}{0.41646} & \multicolumn{2}{c}{0.48507} \\
Post. prob. causal effect: & \multicolumn{2}{c}{58\%} & \multicolumn{2}{c}{51\%} \\ \hline\hline
 & \multicolumn{4}{c}{\textbf{Daily crime counts up to March 28\textsuperscript{th}}} \\
 & \multicolumn{2}{c}{\textbf{Univariate}} & \multicolumn{2}{c}{\textbf{With Cov.}} \\
 & \textbf{Avg.} & \textbf{Cum.} & \textbf{Avg.} & \textbf{Cum.} \\ \hline
Actual & 46 & 1140 & 46 & 1140 \\
Prediction (S.D.) & 45 (2) & 1123 (49) & 46 (1.7) & 1141 (43.3) \\
95\% C.I. & {[}41, 49{]} & {[}1032, 1214{]} & {[}42, 49{]} & {[}1054, 1225{]} \\ \hline
Absolute Effect (S.D.) & 0.67 (2) & 16.84 (49) & -0.055 (1.7) & -1.369 (43.3) \\
95\% C.I. & {[}-2.9, 4.3{]} & {[}-73.6, 107.8{]} & {[}-3.4, 3.4{]} & {[}-85.0, 86.2{]} \\ \hline
Relative Effect (S.D.) & 1.5\% (4.4\%) & 1.5\% (4.4\%) & -0.12\% (3.8\%) & -0.12\% (3.8\%) \\
95\% C.I. & {[}-6.5\%, 9.6\%{]} & {[}-6.5\%, 9.6\%{]} & {[}-7.4\%, 7.5\%{]} & {[}-7.4\%, 7.5\%{]} \\ \hline
Post. tail-area prob. p: & \multicolumn{2}{c}{0.37151} & \multicolumn{2}{c}{0.49896} \\
Post. prob. causal effect: & \multicolumn{2}{c}{63\%} & \multicolumn{2}{c}{50\%} \\ \hline
\end{tabular}
\caption{Causal Impact Analysis - Stolen Vehicles}
\label{vehicles}
\end{table}

\begin{table}[!hbt]
\centering
\footnotesize
\begin{tabular}{lcccc}
\hline
 & \multicolumn{4}{c}{\textbf{Daily crime counts up to March 16\textsuperscript{th}}} \\ 
\textbf{} & \multicolumn{2}{c}{\textbf{Univariate}} & \multicolumn{2}{c}{\textbf{With Cov.}} \\ 
 & \textbf{Avg.} & \textbf{Cum.} & \textbf{Avg.} & \textbf{Cum.} \\ \hline
Actual & 0.62 & 8.00 & 0.62 & 8.00 \\
Prediction (S.D.) & 0.72 (0.26) & 9.38 (3.36) & 0.69 (0.25) & 8.94 (3.28) \\
95\% C.I. & {[}0.21, 1.3{]} & {[}2.69, 16.3{]} & {[}0.21, 1.2{]} & {[}2.73, 15.5{]} \\ \hline
Absolute Effect (S.D.) & -0.11 (0.26) & -1.38 (3.36) & -0.072 (0.25) & -0.935 (3.28) \\
95\% C.I. & {[}-0.64, 0.41{]} & {[}-8.28, 5.31{]} & {[}-0.58, 0.41{]} & {[}-7.48, 5.27{]} \\ \hline
Relative Effect (S.D.) & -15\% (36\%) & -15\% (36\%) & -10\% (37\%) & -10\% (37\%) \\
95\% C.I. & {[}-88\%, 57\%{]} & {[}-88\%, 57\%{]} & {[}-84\%, 59\%{]} & {[}-84\%, 59\%{]} \\ \hline
Post. tail-area prob. p: & \multicolumn{2}{c}{0.326} & \multicolumn{2}{c}{0.37437} \\
Post. prob. causal effect: & \multicolumn{2}{c}{67\%} & \multicolumn{2}{c}{63\%} \\ \hline\hline
 & \multicolumn{4}{c}{\textbf{Daily crime counts up to March 28\textsuperscript{th}}} \\
 & \multicolumn{2}{c}{\textbf{Univariate}} & \multicolumn{2}{c}{\textbf{With Cov.}} \\
 & \textbf{Avg.} & \textbf{Cum.} & \textbf{Avg.} & \textbf{Cum.} \\ \hline
Actual & 0.52 & 13.00 & 0.52 & 13.00 \\
Prediction (S.D.) & 0.72 (0.19) & 18.11 (4.83) & 0.69 (0.19) & 17.14 (4.84) \\
95\% C.I. & {[}0.34, 1.1{]} & {[}8.39, 27.3{]} & {[}0.31, 1{]} & {[}7.71, 26{]} \\ \hline
Absolute Effect (S.D.) & -0.2 (0.19) & -5.1 (4.83) & -0.17 (0.19) & -4.14 (4.84) \\
95\% C.I. & {[}-0.57, 0.18{]} & {[}-14.34, 4.61{]} & {[}-0.52, 0.21{]} & {[}-13.04, 5.29{]} \\ \hline
Relative Effect (S.D.) & -28\% (27\%) & -28\% (27\%) & -24\% (28\%) & -24\% (28\%) \\
95\% C.I. & {[}-79\%, 25\%{]} & {[}-79\%, 25\%{]} & {[}-76\%, 31\%{]} & {[}-76\%, 31\%{]} \\ \hline
Post. tail-area prob. p: & \multicolumn{2}{c}{0.13814} & \multicolumn{2}{c}{0.2012} \\
Post. prob. causal effect: & \multicolumn{2}{c}{86\%} & \multicolumn{2}{c}{80\%} \\ \hline
\end{tabular}
\caption{Causal Impact Analysis - Homicides}
\label{homicides}
\end{table}

\begin{table}[!hbt]
\centering
\footnotesize
\begin{tabular}{lcccc}
\hline
 & \multicolumn{4}{c}{\textbf{Daily crime counts up to March 16\textsuperscript{th}}} \\
 & \multicolumn{2}{c}{\textbf{Univariate}} & \multicolumn{2}{c}{\textbf{With Cov.}} \\ \hline
 & \textbf{Avg.} & \textbf{Cum.} & \textbf{Avg.} & \textbf{Cum.} \\ \hline
Actual & 558 & 6700 & 558 & 6700 \\
Prediction (S.D.) & 592 (14) & 7098 (165) & 590 (13) & 7086 (152) \\
95\% C.I. & {[}567, 620{]} & {[}6805, 7441{]} & {[}564, 615{]} & {[}6772, 7376{]} \\ \hline
Absolute Effect (S.D.) & -33 (14) & -398 (165) & -32 (13) & -386 (152) \\
95\% C.I. & {[}-62, -8.8{]} & {[}-741, -105.1{]} & {[}-56, -6{]} & {[}-676, -72{]} \\ \hline
Relative Effect (S.D.) & -5.6\% (2.3\%) & -5.6\% (2.3\%) & -5.4\% (2.1\%) & -5.4\% (2.1\%) \\
95\% C.I. & {[}-10\%, -1.5\%{]} & {[}-10\%, -1.5\%{]} & {[}-9.5\%, -1\%{]} & {[}-9.5\%, -1\%{]} \\ \hline
Post. tail-area prob. p: & \multicolumn{2}{c}{0.00555} & \multicolumn{2}{c}{0.01493} \\
Post. prob. causal effect: & \multicolumn{2}{c}{99.44\%} & \multicolumn{2}{c}{98.51\%} \\\hline\hline
 & \multicolumn{4}{c}{\textbf{Daily crime counts up to March 28\textsuperscript{th}}} \\
 & \multicolumn{2}{c}{\textbf{Univariate}} & \multicolumn{2}{c}{\textbf{With Cov.}} \\
 & \textbf{Avg.} & \textbf{Cum.} & \textbf{Avg.} & \textbf{Cum.} \\ \hline
Actual & 513 & 12829 & 513 & 12829 \\
Prediction (S.D.) & 600 (12) & 1123 (49) & 596 (9) & 14899 (224) \\
95\% C.I. & {[}580, 624{]} & {[}14490, 15600{]} & {[}580, 614{]} & {[}14500, 15341{]} \\ \hline
Absolute Effect (S.D.) & -87 (12) & -2182 (300) & -83 (9) & -2070 (224) \\
95\% C.I. & {[}-111, -66{]} & {[}-2771, -1661{]} & {[}-100, -67{]} & {[}-2512, -1671{]} \\ \hline
Relative Effect (S.D.) & -15\% (2\%) & -15\% (2\%) & -14\% (1.5\%) & -14\% (1.5\%) \\
95\% C.I. & {[}-18\%, -11\%{]} & {[}-18\%, -11\%{]} & {[}-17\%, -11\%{]} & {[}-17\%, -11\%{]} \\ \hline
Post. tail-area prob. p: & \multicolumn{2}{c}{0.00279} & \multicolumn{2}{c}{0.00208} \\
Post. prob. causal effect: & \multicolumn{2}{c}{99.72\%} & \multicolumn{2}{c}{99.79\%} \\ \hline
\end{tabular}
\caption{Causal Impact Analysis - Overall Crimes}
\label{all}
\end{table}

\end{document}